\documentclass[acmsmall]{acmart}
\AtBeginDocument{%
  }

\setcopyright{cc}
\setcctype{by}
\acmJournal{PACMSE}
\acmYear{2026} \acmVolume{3} \acmNumber{ISSTA} \acmArticle{ISSTA093}
\acmMonth{10} \acmDOI{10.1145/3832184}

\usepackage{multirow}
\usepackage{array}
\usepackage{cleveref}
\usepackage{csquotes}
\usepackage{siunitx}

\begin{document}

\title{Evaluating the Impact of Explainable AI on Trust in AI-Assisted Code Review}

\author{Zhenhan Gao}
\email{zhenhan.gao@tum.de}
\orcid{0009-0000-3305-6770}
\affiliation{%
  \institution{Technical University of Munich}
  \city{Heilbronn}
  \country{Germany}
}

\author{Marvin Muñoz Barón}
\email{marvin.munoz-baron@tum.de}
\orcid{0000-0001-5991-3072}
\affiliation{%
  \institution{Technical University of Munich}
  \city{Heilbronn}
  \country{Germany}
}

\author{Umm-e Habiba}
\email{umme.habiba@tum.de}
\orcid{0000-0001-8953-9624}
\affiliation{%
  \institution{Technical University of Munich}
  \city{Heilbronn}
  \country{Germany}
}

\author{Daniel Graziotin}
\orcid{0000-0002-9107-7681}
\affiliation{%
  \institution{University of Hohenheim}
  \city{Hohenheim}
  \country{Germany}}
\email{graziotin@uni-hohenheim.de}

\author{Stefan Wagner}
\email{stefan.wagner@tum.de}
\orcid{0000-0002-5256-8429}
\affiliation{%
  \institution{Technical University of Munich}
  \city{Heilbronn}
  \country{Germany}
}

\begin{abstract}
\textbf{Background:} The use of large language models (LLMs) for automated code review has brought significant change to a time-consuming part of software engineering.
Prior work has shown that LLM-based code tools can improve code quality and enable more robust software development processes.
As the tools get more powerful, the explanations behind their decisions remain hard to understand.
Developers struggle to assess the validity of LLM-generated code reviews, making it difficult to gauge how much trust they should place in them.
While the application of automated code review with LLMs has been extensively investigated, the inclusion of Explainable AI (XAI) for transparency in code reviews and its impact on trust are yet to be explored.
\\
\textbf{Objective:} We aim to address this research gap by studying the influence of XAI on the trust of software developers in AI-assisted code reviews. \\
\textbf{Method:} We conducted a within-subjects user study with 34 participants from diverse programming backgrounds, comparing three experimental LLM-based automated code review systems with varying levels of XAI support: Condition A (detailed explanation and review feedback), Condition B (review feedback only), and Condition C (no explanations).
Participants were shown a series of real-world code change requests along with the AI-generated code reviews.
During the study, we measured trust perceptions for each system using a questionnaire, agreement with the AI recommendation, the reasoning for accepting or rejecting the code change, and the time taken to review the code change.
 \\
\textbf{Results:} Our quantitative results show that the level of explanation significantly influences both the level of trust of software developers and their agreement with AI recommendations, but in different ways.
Full explanations (Condition A) yield the highest perceived trust (M = 3.99/5) but not the highest agreement with AI recommendations, whereas moderate explanations (Condition B) achieve the highest agreement with AI (89.22\%).
This could suggest that more explanations prompt developers to question AI recommendations more frequently.
In contrast, providing no explanations (Condition C) results in the lowest levels of trust and agreement.
We also find that the level of explanation did not significantly impact the time taken to accept or reject a code change.
Across all conditions, the most commonly cited reasons for code change decisions were changes in code readability and the correctness of the implementation.\\
\textbf{Conclusion:} Overall, these findings indicate that incorporating XAI into the code review process significantly changes the trust perceptions and agreement with AI recommendations for software developers.
These results provide insights for the design and evaluation of trustworthy AI-based code review systems, and support researchers in the design of studies on the human factors of AI-assisted software development.

\end{abstract}



\maketitle
\section{Introduction}
Code reviews are a well-established standard in modern software engineering (SE).
The most recent evolution of this process has been the augmentation of traditional reviews performed by humans with large language model (LLM) generated feedback~\cite{pornprasit2024fine, Aelsteinsson2025Rethinking}.
Developers use LLMs to review their code, identify bugs, and refactor code to improve its quality.
In some cases, these AI (artificial intelligence) assistants are directly integrated into continuous integration and deployment (CI/CD) pipelines, where code changes are automatically reviewed before being integrated into the main code base~\cite{sun2025bitsai}.
Adding LLMs to the review pipeline may have positive impacts on the reviewing experience of developers, and AI-led code reviews may even be preferred to human-led ones~\cite{Aelsteinsson2025Rethinking}.
Automated code review has also been shown to improve software code quality~\cite{cihan2025automated}.

While AI shows much promise for the code review process, it is not without its shortcomings.
The tendency of LLMs to hallucinate additional content leads to faulty reviews, unnecessary corrections, and irrelevant comments~\cite{cihan2025automated}.
A longstanding issue with AI based on machine learning is the lack of transparency into its decision-making~\cite{10.3389/fcomp.2023.1096257}.
As the adoption of AI becomes more ubiquitous, the lack of transparency becomes even more pronounced.
Moreover, without an explanation, software engineers struggle to gauge the trustworthiness of an AI system or its output.
In an ideal interaction between a human and an AI system, a user has \textit{warranted} or \textit{calibrated} trust~\cite{jacovi2021formalizing, baltes2025rethinking} in the system.
They can clearly understand the AI system's capabilities, shortcomings, and decision-making process.
One approach to foster calibrated trust is Explainable AI (XAI).
XAI aims to design systems that can ``explain their rationale to a human user, characterize their strengths and weaknesses, and convey an understanding of how they will behave in the future''~\cite{gunning2019darpa}.
The concept of trust is at the forefront of technology adoption, acting as a mediating factor in the proliferation of new tools~\cite{widder2021trust}.
The most prevalent reason individuals seek assistance from humans rather than AI is a lack of trust in the AI’s responses~\cite{stackoverflow2025survey}.
Calls for a stronger focus on investigating trust-related factors in AI adoption are common~\cite{widder2021trust, lambiase2025investigating}.
Despite these frequent calls, trust is often overlooked in research on LLMs for SE, with a focus on improving model performance rather than increasing transparency~\cite {baltes2025rethinking}.
Addressing the trust gap in software engineering research, we explore XAI methods to promote calibrated trust in interactions with AI-supported software tools.

In this work, we aim to investigate the impact of XAI on AI-generated code reviews and how it relates to the acceptance behavior and trust perceptions by software engineers.
Specifically, we conducted a user study to compare acceptance of code changes and software engineers' trust ratings for AI-generated code reviews across varying levels of explanation.
We compared three systems, resulting in three experimental conditions: Condition A, which provided an AI recommendation with full explanations; Condition B, which provided an AI recommendation with some explanations; and Condition C, which provided an AI recommendation with no explanations.
This allowed us to see the difference between a system with XAI and one without XAI, and to inform the design process for future XAI-based systems regarding the length and depth of XAI explanations.
Each participant was shown a series of code changes of varying complexity, along with the corresponding reviews generated by the three systems.
They then indicated whether they would accept or reject the code change, their reasoning for doing so, and rated their level of trust in each system.
Our research methodology was guided by the following research questions.
\paragraph{\textbf{RQ1: How does the use of XAI methods influence the trust of software engineers in AI-assisted code reviews?}}
  AI-generated suggestions may appear reasonable while still being incorrect, and the increasing complexity of modern models makes their behavior difficult to interpret, complicating developers’ trust in AI-assisted tools. 
  XAI methods may shed light on the reasoning behind AI suggestions and influence how developers form trust. 
  RQ1, therefore, investigates how the presence of XAI influences software engineers’ trust in AI-assisted code reviews.
  To address this question, we analyze participants’ trust questionnaire responses collected under each experimental condition in the user study. Each condition corresponds to a set of experimental samples displayed with Condition A, B, or C.
    We formulate the following hypotheses for the analysis of \textbf{RQ1}:
    \begin{itemize}
    \item Null Hypothesis ($H_0$): There is no difference in the mean trust scores of participants among the three conditions.
    \item Alternative Hypothesis ($H_1$): There is a significant difference in the mean trust scores of participants among the three conditions.
    \end{itemize}
    
\paragraph{\textbf{RQ2: How does the level of explanation affect developers’ agreement with AI recommendations?}}
While XAI may affect how developers perceive and trust AI-assisted tools, different levels of explanation may also shape their decision-making. 
Specifically, richer explanations can increase the persuasive power of AI recommendations, but they may also prompt deeper scrutiny and alternative interpretations, potentially leading to disagreement.
Therefore, RQ2 focuses on how varying levels of explanation influence developers’ willingness to align with AI suggestions during code review. To address this question, we compare participants’ agreement with the AI’s recommendations across different conditions.

    We formulate the following hypotheses for the analysis:
    \begin{itemize}
      \item Null Hypothesis ($H_0$): There is no difference in agreement scores of participants among the three conditions.
      \item Alternative Hypothesis ($H_1$): There is a significant difference in agreement scores of participants among the three conditions.
    \end{itemize}
    
\paragraph{\textbf{RQ3: What is the impact of XAI on the time taken to complete the review tasks?}} RQ3 examines whether different XAI formats influence how long developers need to complete AI-assisted code review tasks. By changing how much context and justification are provided, explanations may either streamline decision making or add extra cognitive effort that slows participants down.

    We formulate the following hypotheses for the analysis:
    \begin{itemize}
      \item Null Hypothesis ($H_0$): There is no difference in completion time of participants among the three conditions.
      \item Alternative Hypothesis ($H_1$): There is a significant difference in completion time of participants among the three conditions.
    \end{itemize}

\paragraph{\textbf{RQ4: What reasons do software engineers provide for accepting or rejecting AI code review?}} 
While RQ1, RQ2, and RQ3 indicate whether explanation formats affect trust and task outcomes, they do not reveal the underlying rationale behind participants' decisions. 
We therefore also analyze the reasons participants provided for their acceptance or rejection decisions for the code changes.
As RQ4 is exploratory, we do not formulate an explicit hypothesis.

\paragraph{Results} The results indicate that the level of explanation for AI-generated code reviews significantly influences the trust perceptions of developers.
Systems with more extensive explanations appear more trustworthy to developers.
However, we did not find the same linear relationship regarding the acceptance of AI recommendations; more extensive explanations did not always lead to greater acceptance.
In general, participants followed the AI recommendations most often when presented with a system with only some explanations (B) rather than full explanations (A).
We found differences in the time taken to come to a decision on a code change, with the system with full explanations (A) requiring the most time to review.
However, the time differences were not statistically significant.
Finally, the results of the qualitative coding of participant responses show that the most common reason provided for all three systems was changes to code readability and the correctness of the results.
The system with no explanations (C) was the only one in which participants never explicitly stated agreement with the AI, although they often followed its recommendations.

\paragraph{Paper Structure} The paper is structured as follows:
\Cref{sec:2-background} gives a brief overview of related studies and literature.
\Cref{sec:3-methodology} outlines the step-by-step methodology used to answer the research questions.
In \Cref{sec:results-discussion}, we present the results of our research and discuss them in the context of the wider research area and their implications for research and industry.
\Cref{sec:5-conclusion} provides a summary of the presented work, gives a short outlook for future work, and presents a final conclusion.

\section{Background}
\label{sec:2-background}

\subsection{AI for Code Review} \label{sec:AI4Code}
Since the introduction of the Transformer architecture~\cite{vaswani2017attention}, generative AI (GenAI) has achieved a strong performance in natural language and software development tasks~\cite{fan2023large}, including code generation, testing, and repair. 
Recent models have been further specialized to support a wider range of code-related tasks. 
As AI-assisted code generation tools become increasingly integrated into software development workflows, the task of reviewing code is expected to play an even more important role for developers.
Code review is a collaborative process in which developers evaluate code changes prior to integration to ensure quality, correctness, and maintainability.
It typically involves one or more reviewers who examine the changes and provide feedback to improve the code's reliability and overall quality.
However, code review is time-consuming.
According to Bosu and Carver, developers spend an average of more than 6 hours per week conducting code reviews~\cite{bosu2013impact}. 

Since code review is time-consuming and labor-intensive, prior work has explored automating it using Deep Learning (DL) techniques.
Shi et al.~\cite{shi2019automatic} proposed DACE, which combines CNNs, LSTMs, and a Pairwise Autoencoder to model contextual and revision features of code changes, achieving strong performance across six open-source projects.
Hellendoorn et al.~\cite{hellendoorn2021towards} investigated predicting whether code diffs require review comments, showing that even this fundamental task remains challenging.
Li et al.~\cite{li2019deepreview} introduced DeepReview, an end-to-end DL model that formulates code review as a multi-instance learning problem and predicts approval decisions based on raw code changes and descriptions.

With the emergence of Large Language Models (LLMs), recent studies have examined their potential for code review automation.
Several studies analyze and compare the performance of various LLMs in code-review related tasks~\cite{cihan2025evaluating, watanabe2024use, Aelsteinsson2025Rethinking}.
Cihan et al.~\cite{cihan2025evaluating} further compared GPT-4o and Gemini 2.0 Flash, showing that GPT-4o achieves higher accuracy in both review decision correctness and bug-fixing tasks.
Pornprasit and Tantithamthavorn~\cite{pornprasit2024fine} compared the techniques of fine-tuning and prompt engineering to optimize LLMs for code review tasks.
They generally recommend fine-tuning for best performance when sufficient training data is available and few-shot learning in other cases.
In addition, Watanabe et al.~\cite{watanabe2024use} analyzed 229 review comments from 205 pull requests and found that reviewers often use suggestions generated by ChatGPT as supporting evidence during code review. They also observed that the majority of users have a positive attitude towards ChatGPT, with only 30.7\% of reactions to its answers being negative.
A\dh alsteinsson et al.~\cite{Aelsteinsson2025Rethinking} conducted an industrial case study comparing two different LLM-assisted code review tools.
They found that AI-led pull requests are generally preferred over those initiated by humans, with reviewers' familiarity with the codebase and the severity of the pull request influencing this preference.
In their field study, Sun et al.~\cite{sun2025bitsai} tested an LLM-based system for automated code reviews in a large-scale industrial environment.
They report a successful tool introduction, citing high adoption rates among developers.
The current state of research focuses on improving the accuracy and performance of machine learning models for code review and on evaluating automated code review systems in practice.
However, there is a distinct lack of research on increasing the transparency of AI-generated reviews and on designing trustworthy AI systems for automated code review.

\subsection{Explainable AI} \label{sec:Explainable-AI}
Trust is at the core of the many discussions surrounding AI, with many studies stressing its importance in human-AI interactions~\cite{widder2021trust, lambiase2025investigating}.
In general, we aim to achieve an interaction with calibrated trust~\cite{baltes2025rethinking, jacovi2021formalizing}, in which a user neither mistrusts a system without good reason (undertrust) nor places too much trust in a system (overtrust).
Defining at what point an interaction can be considered to have occurred with calibrated trust remains elusive, as measuring human trust is a difficult endeavor.
The most common direct indicator of trust is the perceived trustworthiness of a system, which is often measured by administering a questionnaire after a user has interacted with it~\cite{10.3389/fcomp.2023.1096257}.
Several scales have been proposed to measure users' perceived trust when interacting with automated systems, such as the XAI scale by Hoffman et al.~\cite{10.3389/fcomp.2023.1096257} and the trust scale for automated systems by Jian et al.~\cite{jian2000foundations}.
While trust is important for the adoption of any software tool, it is especially important for those based on machine learning, notably LLMs.
Despite their strong emergent capabilities, LLMs remain susceptible to systematic errors, commonly referred to as hallucinations, which can severely undermine trust, transparency, and accountability in high-stakes applications~\cite{atakishiyev2025explainability}.

As a counterbalance to the black box nature of modern AI-based systems, explainable AI (XAI) has emerged as a key enabler of transparent and trustworthy AI systems, particularly in safety-critical settings~\cite{10.1016/j.knosys.2023.110273, hossain2025explainable, atakishiyev2024explainable}.
Early XAI research primarily relied on explanations as reasoning mechanisms in expert systems or as symbolic abstractions of neural networks; however, the growing complexity of modern deep learning systems has significantly intensified the demand for effective explanation techniques~\cite{atakishiyev2025explainability}.
With the increasing deployment of AI systems in critical domains~\cite{chang2024survey}, these explainability considerations naturally extend to LLMs. 
Recent surveys categorize LLM explainability techniques according to transformer architectures (encoder-only, decoder-only, and encoder--decoder models), examine evaluation protocols for explanation quality, and analyze how explanations are utilized in downstream applications~\cite{Palikhe2025TowardsTA}.
Complementing this architectural perspective, Wu et al.~\cite{Wu2024} argue that traditional XAI paradigms are insufficient for LLMs and propose a shift toward \emph{usable XAI}, where explanations not only expose model behavior but also actively improve LLM performance, productivity, and human--AI interaction. 
Their framework highlights bidirectional opportunities: using XAI to better understand and refine LLMs, and leveraging LLMs themselves to enhance explainability techniques.

Beyond general-purpose language understanding, LLM explainability has also been explored in application-specific contexts.
In recommender systems, for instance, Said~\cite{said2025explaining} provides a focused review of how LLMs are employed to generate natural-language explanations for recommendations.
Despite the growing popularity of LLMs, the review finds that only a small number of studies explicitly address explainable recommendation generation with LLMs, suggesting that this research direction remains nascent.
Nevertheless, these early efforts demonstrate the potential of LLMs to produce more transparent, user-centric explanations, reinforcing the broader role of explainability in fostering trust and accountability across LLM-driven systems.
In the code review domain, the use of XAI is particularly sparse.
Existing work has primarily addressed isolated subtasks, such as toxicity detection~\cite{sarker2023toxispanse} or comment quality classification~\cite{yang2023evacrc}, rather than the full code review process.
This limited adoption is likely due to the composite nature of code review, which involves multiple intertwined subtasks, including bug detection, quality assessment, reasoning about code changes, and decision-making.
As a result, existing XAI techniques do not readily transfer to end-to-end, generative code review scenarios.

Overall, prior work suggests that while XAI techniques are well established for classification tasks, their application to GenAI, especially for code review, remains underexplored~\cite{sun2022investigating}, highlighting a clear gap that our work aims to address.
\section{Methodology}
\label{sec:3-methodology}

\begin{figure}[htbp]
  \centering
  \includegraphics[width=0.6\linewidth]{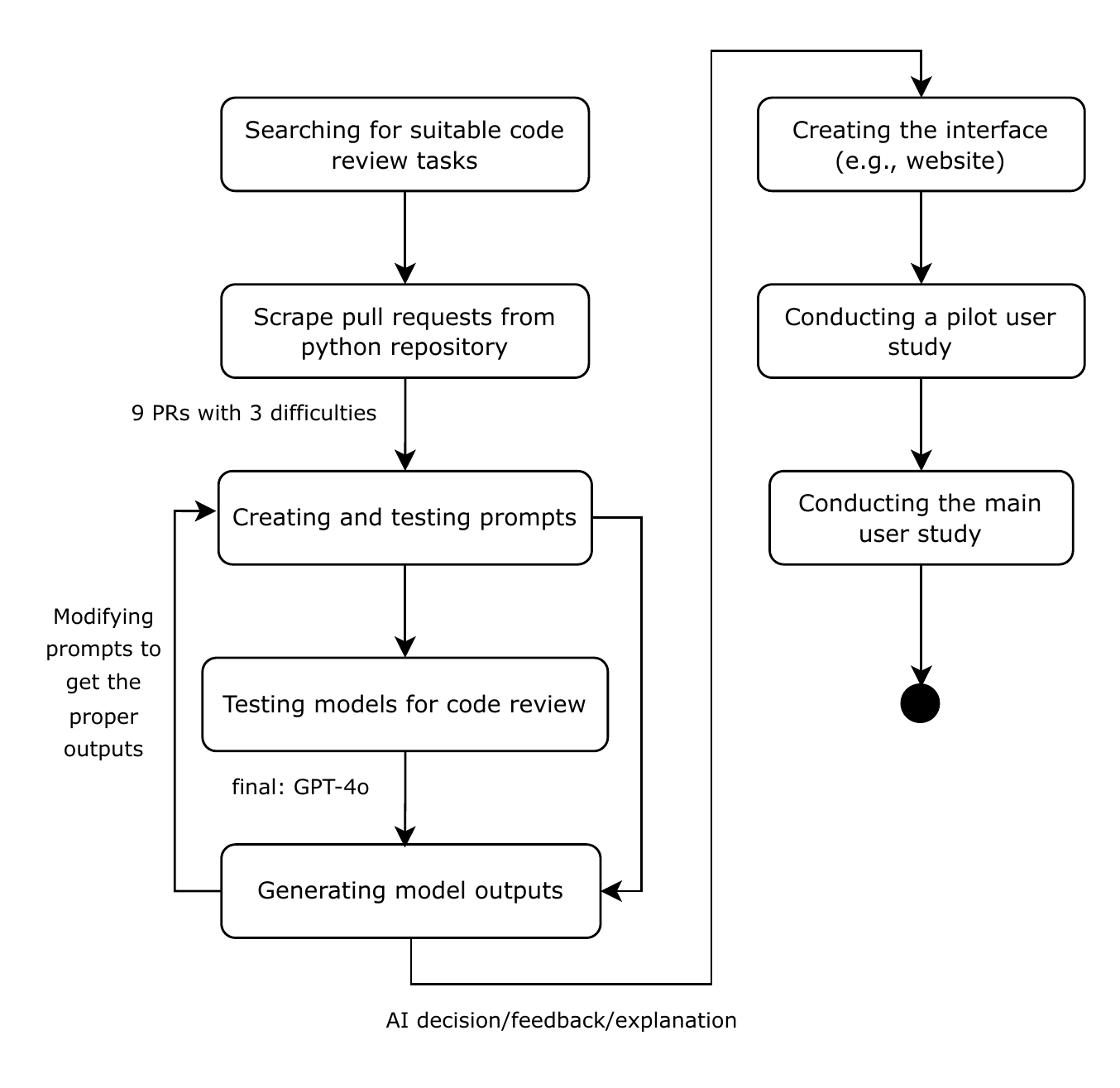}
  \caption{Overview of the research workflow}
  \Description{A workflow diagram illustrating the development process, including data collection, model selection, interface creation, and user study.}
  \label{fig:dev_overview}
\end{figure}

The primary objective of this study is to examine how XAI affects software engineers’ trust and performance for code review.
We conducted a controlled user study in which participants completed the same set of code review tasks under three experimental conditions:
Condition A, where participants see detailed inline explanations, the review comments, and the AI recommendation regarding acceptance or rejection of the code change.
The inline explanations are short, localized rationales embedded directly into the diff, highlighting the specific code fragments that most strongly influence the AI's decision (i.e., which parts of the code play a critical role).
The review comment is presented separately from the code as a list of feedback suggestions.
The second is Condition B, in which participants see the review comments together with the binary AI recommendation.
Condition C is the third condition, where only the binary AI recommendation is shown.
To support the study, we implement a web-based code review platform and curate nine predefined real-world pull requests (PRs) from a Python repository as experimental materials.
Across conditions, we record the participants' decisions and completion time, and we measure trust via a questionnaire administered immediately after each condition.
Based on the results, we provide insights into how different explanation levels influence users’ trust, and overall performance in AI-assisted code review.
Each step is briefly illustrated in Figure~\ref{fig:dev_overview}. 
In the remaining section, we describe each of these steps in detail, first introducing the automated code review system used in the study in~\Cref{subsec:xai-system}, followed by the details of the user study itself in~\Cref{subsec:user-study}. 

\subsection{XAI System for Code Reviews}
\label{subsec:xai-system}

\subsubsection{Architecture}
The generation of the code reviews follows a two-phase XAI pipeline.
In Phase~1, an LLM receives the proposed code change, including the added and removed lines, along with a natural-language prompt instructing it to review the changes, provide a recommendation to \textbf{accept} or \textbf{reject} the change, and list \textbf{reasons} for the recommendation.
Phase~2 builds on the Phase~1 output. The LLM is prompted to further \textbf{explain} the generated feedback and \textbf{highlight} the code sections that are relevant to the individual feedback suggestions. This output is then presented as inline explanations and code highlighting, linking the review feedback to specific elements of the code change, such as function names, comments, and logic structures.

The three user study conditions correspond to cumulative levels of this pipeline rather than independently designed feedback formats.
Condition C exposes only the Phase 1 recommendation. Condition B exposes the complete Phase 1 feedback, including both the recommendation and its reasons, but does not include any Phase~2 output.
Condition A includes the same Phase 1 feedback as Condition B and additionally exposes the Phase~2 output through inline explanations and highlighted code sections. Thus, Condition A is not simply a more detailed version of Condition B.
Rather, it supplements the Phase 1 recommendation and reasons with a distinct Phase~2 explanation layer that grounds individual feedback suggestions in the relevant code sections.

\paragraph{Models}\label{sec:models}
Both steps were performed with ChatGPT using the GPT-4o model.
When designing the system, we considered alternative models for generating the code reviews for our user study.
We examined CodeReviewer, a publicly available model on HuggingFace~\cite{codereviewer2022huggingface}, but decided against using it due to its limited output quality and limited customizability for our use case.
We also considered more traditional XAI methods, such as SHAP or LIME, but these methods are mainly applicable to smaller, accessible models and produce outputs that are difficult for developers to interpret in code-review tasks. 
Since LLM-generated explanations may be post-hoc rationalizations, we prompted the model to provide feedback before giving its final recommendation. This encourages consistency between the explanation and the decision, but does not guarantee faithfulness.

In our final system, both the code review and explanations are generated with ChatGPT using the GPT-4o model.
This choice is motivated by prior studies by Tufano et al. and Cihan et al.~\cite{tufano2024code, cihan2025evaluating} that showed that, on code-related tasks, ChatGPT performs slightly better than specialized code review models and Gemini.
GPT-4o produces a well-structured review that matches our prompt requirements, including clearly organized feedback with descriptive headings and a final accept/reject decision.
During generation, we set the temperature to zero to reduce the variance in the generated output.

\begin{figure}[htbp]
  \centering
  \includegraphics[width=0.9\linewidth]{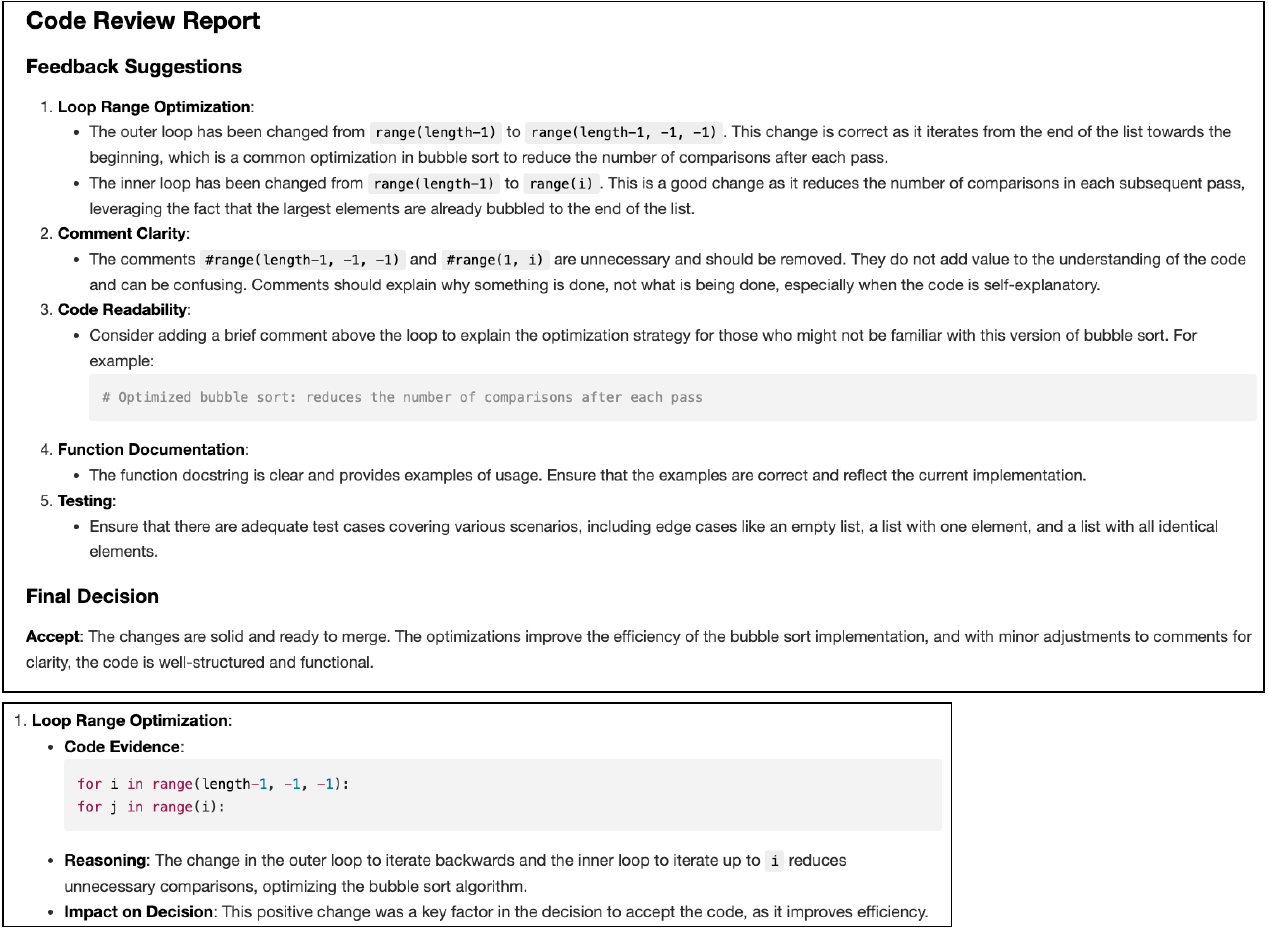}
  \caption{Outputs generated by the XAI system for a single pull request.}
  \Description{The figure shows the structured review report generated by the first-phase model, including feedback suggestions and the final accept/reject decision, together with a second-phase example explanation that links specific code changes to the model’s reasoning and decision rationale.}
  \label{fig:gpt4o_output}
\end{figure}

\paragraph{Prompts} The full prompts used in both steps are reported in the replication package.
The prompts were manually designed through an iterative process of altering the prompt and evaluating the output.
We randomly sampled code changes from the \texttt{TheAlgorithms/Python} repository~\cite{thealgorithms}, which subsequently served as experimental samples in our user study. The rationale for selecting this repository is described in \Cref{subsec:user-study}. To validate the output, we verified the presence of the requested review elements, i.e., accept/reject recommendations, reasons, explanations, and highlighted code sections, and we manually gauged the accuracy of the contents of those elements.
Once the selected prompts passed validation without fail for at least 9 code changes, we considered them sufficiently performant to use for the final XAI system.
Figure~\ref{fig:gpt4o_output} shows the outputs for a single code change for the initial prompt in phase 1 and the follow-up prompt in phase 2.

\subsection{Experimental Setup}
\label{subsec:user-study}

\subsubsection{Experimental Samples}\label{sec:tasks}
As the experimental samples for our study, we selected a set of pull requests (PRs) from real-world projects.
When selecting the PRs, we defined the following six criteria to ensure suitability for our user study:

\begin{enumerate}
    \item \label{crit:language} The code must be written in a widely used programming language.
    \item \label{crit:nontrivial} The changes must be non-trivial (e.g., not limited to documentation or formatting).
    \item \label{crit:diff} Each PR must contain a clear code diff, showing which lines were added or removed.
    \item \label{crit:decision} The final decision on the PR, whether it is accepted or rejected, must be available.
    \item \label{crit:selfcontained} The PR should be understandable based solely on its own context and basic programming knowledge.
    \item \label{crit:manageable} The overall size of the PR should be manageable, allowing participants to complete the review within a reasonable timeframe during the experiment.
\end{enumerate}

We found that existing representative datasets for code review tasks are mainly designed for model training and often contain highly complex code.
This makes them unsuitable for our study, as our tasks require a controlled level of complexity for participants.
So we opt to construct a custom dataset by manually selecting PRs from open-source projects on GitHub.
After reviewing the top 100 most-starred Python repositories~\cite{githubranking}, we chose the repository TheAlgorithms/Python~\cite{thealgorithms}.
Approximately 99\% of the code is written in the widely used language Python, which satisfies criterion~\ref{crit:language}.
PRs were collected via a Python-based crawler using the GitHub REST API~\cite{githubrestapi}.
We select PRs with diverse, non-trivial change types to satisfy criterion~\ref{crit:nontrivial}.
For each PR, we extract metadata and change statistics, and use the merge outcome (accepted/rejected) as ground truth, satisfying criterion~\ref{crit:decision}.
All tasks are based on the raw diff files retrieved from GitHub, which explicitly include the modified code and line-level changes, satisfying criterion~\ref{crit:diff}.
To ensure independence and readability, we focus on PRs that modify basic algorithmic implementations, typically within a single class and without dependencies on additional files.
We further restrict PRs to fewer than 55 changed lines to keep tasks manageable, satisfying the criteria \ref{crit:selfcontained} and \ref{crit:manageable}.
Task difficulty is determined using both change size (as a coarse proxy) and a qualitative assessment of semantic complexity (e.g., typos/renaming vs.\ algorithmic logic changes).
The final set of pull requests is summarized in Table~\ref{tab:pr-selection} and is available in the replication package.

\begin{table}[htbp]
\centering
\caption{Selected PRs categorized by difficulty and merge decision}
\label{tab:pr-selection}
\small
\begin{tabular}{clll}
\textbf{PR ID} & \textbf{File name} & \textbf{Difficulty} & \textbf{Decision} \\ \hline
10899 & equilibrium\_index\_in\_array.py & easy   & accept \\
11366 & and\_gate.py                     & easy   & reject \\
4485  & average\_mean.py                 & easy   & accept \\
1017  & find\_lcm.py                     & medium & reject \\
7429  & factors.py                       & medium & accept \\
1011  & naive\_string\_search.py         & medium & reject \\
435   & bubble\_sort.py                  & hard   & accept \\
1084  & decimal\_to\_binary.py           & hard   & reject \\
1063  & radix\_sort.py                   & hard   & reject 
\end{tabular}

\end{table}

\subsubsection{Experimental Conditions \& Task Assignment}
Each of the nine experimental samples was then fed into our three AI code review systems, with each system implementing one of the three experimental conditions, resulting in 27 unique final samples.
Table~\ref{tab:condition} summarizes the specific AI components included in each condition.
More details about the two phases design in our system can be found in~\Cref{subsec:xai-system}.

\begin{table}[htbp]
\centering
\caption{Overview of Elements included in Each Experimental Condition}
\label{tab:condition}
\small
\begin{tabular}{cccc}
\hline
Condition & AI Recommendation & Review Feedback & Detailed Explanation \\ \hline
A         & \checkmark & \checkmark       & \checkmark     \\
B         & \checkmark & \checkmark       &                \\
C         & \checkmark &                  &                \\
\hline
\end{tabular}
\end{table}

The study was conducted in a within-subjects setup, with each participant seeing all three conditions in a random order.
For each condition, the participants received three samples, one of each difficulty level.
Both the assignment of tasks to conditions and the task order were also randomized.
In total, each participant saw all nine pull requests.
We note that a flaw in the shuffling procedure led to an imbalance in how the task sets were paired with conditions which we describe and discuss in~\Cref{sec:dis-ttv}.

Across the nine pull requests, the first-phase review feedback used in Condition A and B contained 5 to 10 feedback items (M = 7.7) and ranged from 254 to 407 words (M = 320), each item pairing a short descriptive heading with a one- to three-sentence justification.
The issues raised cover documentation, variable naming, code formatting, edge cases and testing, error handling, efficiency, redundant code, type annotations, and correctness.
The second-phase detailed explanations used only in Condition A are longer, ranging from 391 to 589 words (M = 473) and follow a fixed structure for each feedback item, stating the code evidence the item derives from, the reasoning that links that evidence to the feedback, and the impact of the item on the final accept/reject decision.
\Cref{fig:gpt4o_output} illustrates both outputs for a single pull request, and the complete set of pull requests, prompts, and raw generated reviews is available in the replication package~\cite{gao_2026_21457282}.

\subsubsection{Instrument \& Measures}
\begin{figure}[htbp]
  \centering
  \includegraphics[width=0.8\linewidth]{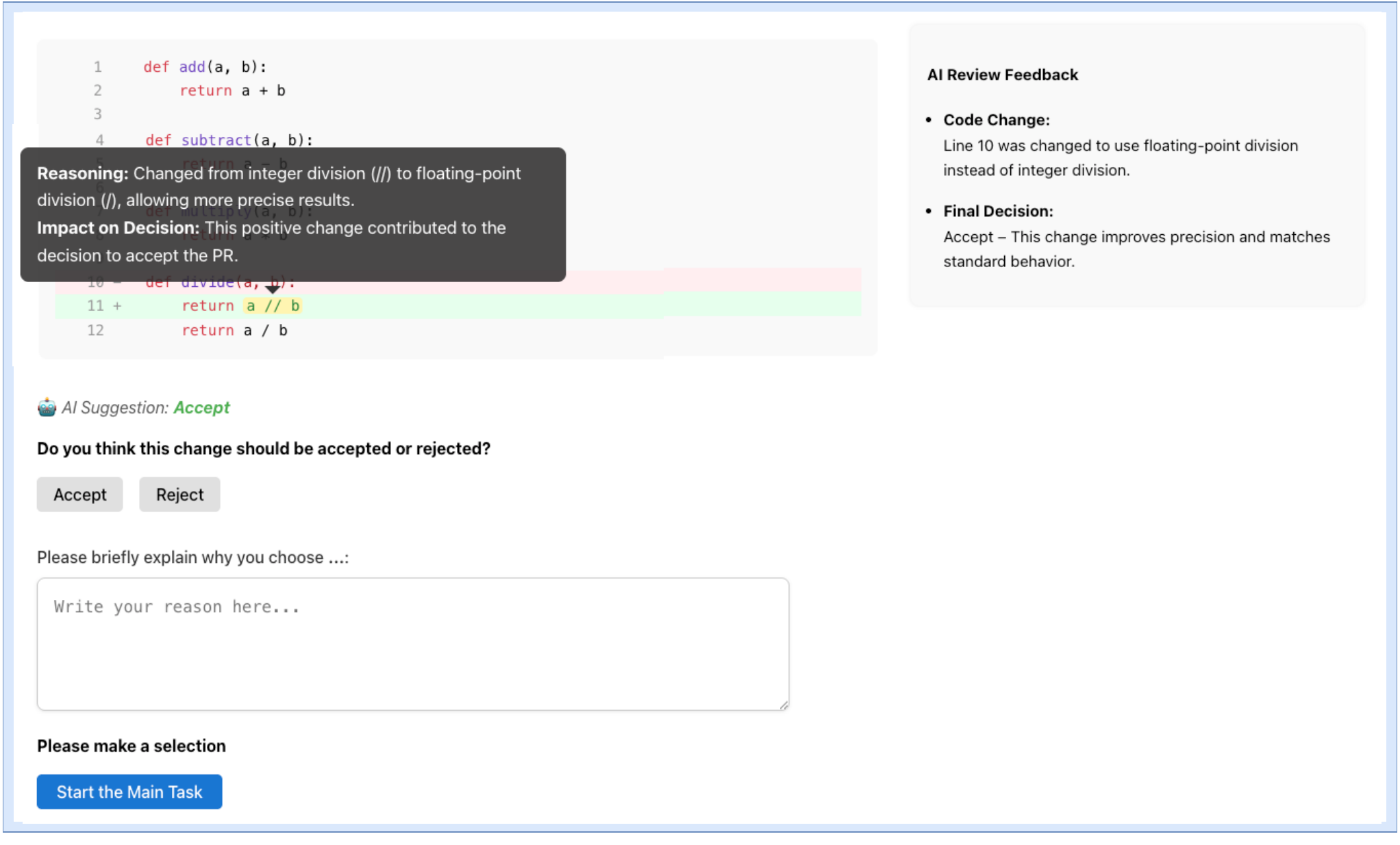}
  \caption{Code Review Task Interface}
  \Description{Example of the code review interface used in the user study.}
  \label{fig:code_review_task_interface}
\end{figure}

The experimental samples were presented to participants via a web application that embeds explanations directly into syntax-highlighted code changes, providing a realistic code-review environment for the user study. 
Figure~\ref{fig:code_review_task_interface} shows the task interface, with the code changes on the left, review comments on the right, and the highlighted detailed explanations, representing Condition A.
In Conditions A and B, the feedback items are listed as review comments in a panel beside the code, in the order the model produced them.
Condition A additionally highlights the code fragments each explanation refers to directly in the diff, with participants revealing the corresponding detailed second-phase explanation by hovering over a highlighted fragment, which keeps the justification anchored to the lines it concerns.
Condition C shows neither a comment panel nor any highlighting, leaving only the binary accept/reject recommendation below the diff.

Upon being presented with a code change, participants are instructed to review it, decide whether to accept or reject it by clicking a button, and provide a reason for their decision in a text box.
We collect both their responses and the time taken to provide them in the web application for each code change.
After the participant has seen all three code changes for an experimental condition, they are presented with a questionnaire about their trust perceptions of the condition that they just interacted with.
We measured trust using the TXAI scale introduced by Hoffman et al.~\cite{10.3389/fcomp.2023.1096257}, excluding the sole reverse-coded item, ``I am wary of the AI.'' Perrig et al.’s psychometric validation found that removing this item improved the CFA model fit (e.g., CFI from .983 to .995 and RMSEA from .056 to .036) and thus recommended this revised version of the scale~\cite{perrig2023trust}.
The seven TXAI items included in our trust questionnaire are reported verbatim, together with their item-level results, in~\Cref{tab:ANOVA_trust_item}.
Additionally, we gathered demographic information, including age, education level, and job role, as well as two factors relevant to task performance: experience with AI tools and Python proficiency.

\subsubsection{Pilot User Study}\label{sec:pilot_user_study}
Before launching the main study, we conducted a pilot study with five participants to assess interface usability, task duration, and task difficulty.
The pilot included two XAI conditions (Condition A and Condition B) and revealed a large variability in completion times (M = \SI{21}{\minute}~\SI{46}{\second}, Md = \SI{13}{\minute}~\SI{17}{\second}, min = \SI{4}{\minute}~\SI{33}{\second}, max = \SI{50}{\minute}~\SI{41}{\second}).
Some participants completed tasks very quickly, suggesting shallow review behavior.
To address this, we added a mandatory reasoning field requiring participants to briefly justify their decisions.
Based on the pilot, we also introduced a third condition (Condition C) that provides only the AI recommendation, without any explanations, to serve as a control condition.

\subsubsection{User Study}\label{sec:user_study}
For our main user study, we determined the required sample size using a power analysis conducted with G*Power 3.1~\cite{faul2007g, gpower}.
Following recommendations for within-subject repeated-measures ANOVA~\cite{langenberg2023tutorial}, we assume a medium effect size (f = 0.25), power = 0.8, and a nonsphericity correction of $\epsilon = 0.75$.
With three conditions per participant and an assumed correlation of 0.5 between repeated measures, the analysis indicates a required sample size of 34 participants.
Based on this result, we recruited participants via Prolific~\cite{prolific} in May 2025, offering a compensation rate of \$7 per hour.

We apply three prescreen criteria during participant recruitment via the Prolific platform: (1) fluency in English, (2) experience with computer programming, and (3) familiarity with the Python programming language.
These filters ensure a suitable participant pool for the study.
In total, N = 34 participants completed the user study as expected, and their data were successfully collected.
We recruited exactly the 34 participants indicated by the a priori power analysis described above; no participants were excluded or filtered out, and all 34 are included in every analysis reported in this paper.
The average age of participants is 31 years (Md = 29.5 years, SD = 10.76 years, min = 21 years, max = 67 years).
The average completion time is approximately \SI{37.9}{\minute} (Md = \SI{35}{\minute}~\SI{32}{\second}, SD = \SI{20}{\minute}~\SI{14}{\second}, min = \SI{4}{\minute}~\SI{10}{\second}, max = \SI{119}{\minute}~\SI{50}{\second}).
On average, participants report an AI experience level of 4.18 (min = 2, max = 5) and a Python programming proficiency of 3.18 (min = 1, max = 5), both on a 5-point scale.
In terms of education, 20 participants hold a Bachelor's degree, 8 a Master's degree, 3 a PhD, and 3 report a high school degree as their highest qualification.
The reported job roles are diverse, including professional software engineers (8), full-stack (5), frontend (2), and backend (1) developers, with the remaining participants being students (8) and participants who selected \enquote{Other} (9).
Because averages can obscure the spread of individual responses, we report the full set of per-participant characteristics in the replication package~\cite{gao_2026_21457282}.
All demographic measures are self-reported through our study questionnaire.

\subsection{Analysis Procedures}

We present the procedures used to analyze the experimental data for each research question.

\paragraph{\textbf{RQ1: How does the use of XAI methods influence the trust of software engineers in AI-assisted code reviews?}}
For each participant and condition, we compute a personal trust score by averaging responses across the seven TXAI items.
We summarize trust per condition using descriptive statistics. To test whether trust differs across the three conditions in our within-subject design, we apply a repeated-measures ANOVA.
We also conduct post hoc two-tailed paired t-tests for the three pairwise comparisons (A-B, A-C, B-C) and apply the Bonferroni correction for multiple comparisons.
In addition, we perform an item-level analysis by computing per-item mean scores for each condition and running separate repeated-measures ANOVAs for each TXAI item (Q1--Q7).
Finally, we report partial eta squared ($\eta_p^2$) as an effect size measure.

\paragraph{\textbf{RQ2: How does the level of explanation affect developers’ agreement with AI recommendations?}}
To assess AI agreement across conditions, we compare each participant’s accept or reject decision for each pull request (PR) with the corresponding AI recommendation.
For each condition, each participant completes three tasks.
We compute an agreement score per participant and condition as the proportion of the three decisions that match the AI recommendation.
To test whether agreement differs across the three conditions in our within-subject design, we apply a repeated-measures ANOVA to participants’ per-condition accuracy scores.
We also conduct post hoc two-tailed paired t-tests for the three pairwise comparisons (A--B, A--C, B--C) and apply the Bonferroni correction for multiple comparisons.

\paragraph{\textbf{RQ3: What is the impact of XAI on the time taken to complete the review tasks?}}
As a performance metric, we measured the task duration for each review task.
For each task, we recorded the duration from the moment the task page is loaded until the participant proceeds to the next page.
For each participant and condition, we compute the average task completion time across the three tasks.
To test whether completion time differs across the three conditions in our within-subject design, we applied a repeated-measures ANOVA on participants' average completion times for each condition.

\paragraph{\textbf{RQ4: What reasons do software engineers provide for accepting or rejecting AI code review?}}
\label{subsec:reason}
Each task includes a mandatory reasoning field requiring participants to briefly justify their accept or reject decisions.
We employed qualitative coding to analyze the participant responses.
Each sample was assigned one or multiple codes, each representing a unique reason for accepting or rejecting a code change.
The coding process was performed by three coders, including two PhD students and one master’s student.
Coding was completed in four rounds.
Initially, a random sample of 10\% of responses was chosen for an explorative open-coding round to build an initial codebook.
Following the open-coding round, an initial codebook was drafted after discussing the independent coding results.
Using the initial codebook, another round of coding was performed, with a new random sample of 10\% of responses.
This second round of coding did not yield reliable results, with inter-rater agreement (Krippendorff's alpha~\cite{krippendorff2018content}) being below 0.6.
The codebook was subsequently refined, and a third round of coding, again with a different random sample of 10\% of responses, was performed.
This refined codebook resulted in a more reliable outcome with an inter-rater agreement of approximately 0.75.
We therefore decided to code the remaining 90\% of samples, including the samples from rounds 1 and 2, using this refined codebook.
All three reviewers were involved in all four coding rounds.
The final hierarchical code book can be found in the replication package.
We compute Krippendorff’s alpha~\cite{krippendorff2018content} on the 306 multi-label reasoning responses coded by three coders. Overall agreement is acceptable ($\alpha = 0.7564$, tentative), indicating reasonable reliability but below the commonly used robustness threshold ($\geq 0.80$).
Overall, Krippendorff’s alpha indicates acceptable agreement, suggesting that the multi-label coding was sufficiently reliable to address the exploratory research question.

\section{Results \& Discussion}
\label{sec:results-discussion}

\subsection{RQ1: How Does the Use of XAI Methods Influence the Trust of Software Engineers in AI-assisted Code Reviews?}

We report participants’ trust scores across the three experimental conditions with different explanation formats, measured using the TXAI scale~\cite{perrig2023trust}.

\paragraph{Descriptive statistics} 
Recall that Condition A provided Phase 1 feedback (recommendation and reasons) together with Phase 2 inline explanations and code highlighting; Condition B provided only the Phase 1 feedback; and Condition C provided only the acceptance or rejection recommendation.
The mean trust score is highest for Condition A (M = 3.99, Md = 4.0, SD = 0.57, min = 3.0, max = 5.0), followed by Condition B (M = 3.77, Md = 3.93, SD = 0.8, min = 1.29, max = 5.0), and lowest for Condition C (M = 3.41, Md = 3.43, SD = 0.81, min = 1.86, max = 5.0), indicating that the format of XAI influences participants’ trust in the AI-assisted code review process.
Condition A shows the least variance between trust ratings among participants, with no significant outliers and a narrow range between quartiles.
Condition B shows a slightly more dispersed score, with a few lower outliers, suggesting that while the majority of participants trust the system, a few expressed significant skepticism.
Condition C shows the lowest median and widest spread, including very low scores, suggesting greater variability in trust and lower overall trust in the system when no or limited XAI is provided.

\paragraph{Statistical test} 
A repeated-measures ANOVA reveals a significant effect of the condition on trust, $F(2, 66) = 8.2497, p = 0.0006 < 0.05$.
We therefore reject $H_0$ and conclude that there is \textbf{a significant difference} in the mean trust scores of the participants across the three conditions.

\paragraph{Post-hoc comparisons}
Post-hoc two-tailed paired t-tests show that trust is higher in A than C, $t(33)=4.65, p=0.0001$, and B is slightly higher than C, $t(33)=2.06, p=0.0475$, while A and B do not differ, $t(33)=1.73, p=0.0936$.
After Bonferroni correction for three comparisons ($\alpha=0.017$), only A--C remains significant ($p_\text{corrected}=0.00015$), A--B ($0.2808$) and B--C ($0.1426$) are not significant (Table~\ref{tab:ttests_trust}).
Overall, full explanations (Condition A) elicit significantly higher trust than no explanations (Condition C).

\begin{table}[htbp]
\centering
\small
\caption{Post-hoc pairwise t-tests on trust scores with Bonferroni correction}
\label{tab:ttests_trust}
\begin{tabular}{cccc}
\toprule
\textbf{Comparison} & \textbf{Raw $p$-value} & \textbf{Corrected $p$-value} & \textbf{Significant (Bonferroni)} \\
\midrule
A vs B & 0.0936 & 0.2808 & No \\
{\color[HTML]{CB0000} \textbf{A vs C}} & {\color[HTML]{CB0000} \textbf{0.0001}} & {\color[HTML]{CB0000} \textbf{0.00015}} & {\color[HTML]{CB0000} \textbf{Yes}}        \\
B vs C & 0.0475 & 0.1426 & No \\
\bottomrule
\end{tabular}
\end{table}

\paragraph{Item-Level Trust Analysis}
Across all seven items, mean scores are consistently highest for Condition A, followed by Condition B and then Condition C.
Item-level tests indicate that Q1 shows the strongest effect of condition, $F(2, 66) = 10.85$, $p = 0.0001$, followed by significant differences for Q3, $F(2, 66) = 6.80$, $p = 0.0021$, Q6, $F(2, 66) = 6.33$, $p = 0.0031$, and Q7, $F(2, 66) = 4.79$, $p = 0.0115$.
In contrast, Q2, Q4, and Q5 do not show statistically significant differences across the three conditions ($p > 0.05$). 
The item texts and detailed results for all items are reported in Table~\ref{tab:ANOVA_trust_item}.

\begin{table}[htbp]
\centering
\small
\setlength{\tabcolsep}{3pt}
\caption{Effect of condition on individual TXAI trust items (Q1--Q7)}
\label{tab:ANOVA_trust_item}
\begin{tabular}{clccc}
\toprule
\textbf{Item} & \textbf{Text} & \textbf{F(2,66)} & \textbf{$p$} & \textbf{Sig.} \\
\midrule
Q1 & I am confident in the [system X]. I feel that it works well. & 10.85 & 0.0001 & {\color[HTML]{CB0000}\textbf{Yes}} \\
Q2 & The outputs of the [system X] are very predictable. & 1.46 & 0.2387 & No \\
Q3 & The [system X] is very reliable. I can count on it to be correct all the time. & 6.80 & 0.0021 & {\color[HTML]{CB0000}\textbf{Yes}} \\
Q4 & I feel safe that when I rely on the [system X] I will get the right answers. & 2.66 & 0.0772 & No \\
Q5 & The [system X] is efficient in that it works very quickly. & 2.30 & 0.1082 & No \\
Q6 & The [system X] can perform the task better than a novice human user. & 6.33 & 0.0031 & {\color[HTML]{CB0000}\textbf{Yes}} \\
Q7 & I like using the [system X] for decision making. & 4.79 & 0.0115 & {\color[HTML]{CB0000}\textbf{Yes}} \\
\bottomrule
\end{tabular}
\end{table}

\paragraph{Effect size} 
After completing the descriptive statistics and the main statistical test, we further compute and report partial eta squared ($\eta_p^2$) as an effect-size measure.
The analysis showed a significant effect of Condition on trust, $F(2, 66) = 8.25, p = 0.0006$, with a moderate-to-large effect ($\eta_p^2 = 0.2$).

\paragraph{Discussion}
Our findings suggest that the format of XAI affects users’ trust in the AI-assisted code review system.
Specifically, the significantly higher trust scores under Condition A indicate that full explanations, which directly align justifications with specific code changes, are perceived as more transparent and reliable.
This may be because such explanations are embedded within the code itself, enabling users to quickly grasp which specific parts of the code influence the AI's decision.
In contrast, Condition B, which provided only some explanations via review comments, was less precise in mapping reasoning to specific lines of code.
Another possible reasoning is that, in real-world scenarios, developers often encounter lengthy or unfocused AI responses that fail to highlight the most relevant information.
Inline explanations with highlighting, by contrast, present model reasoning in a concise and context-aware way, making it easier for users to interpret the AI’s rationale directly within the code and increasing the overall transparency of the AI recommendation.
In contrast, the lower and more variable trust scores in Condition C highlight the uncertainty and skepticism users experience when no explanation is provided.
Condition B, which offered some explanations, also showed higher trust scores than Condition C, which provided no explanations, but the difference was not significant.
This suggests that the placement and contextual integration of explanations may be critical factors.
The significant differences in Q1, Q3, Q6, and Q7 suggest that the conditions mainly changed how positively participants viewed the AI. Depending on the condition, participants differed in how confident they felt in the AI, how reliable they thought it was, whether they believed it could perform better than a novice human reviewer, and whether they would like to use it in decision making. Overall, the explanations seemed to influence whether participants saw the AI as a capable and useful review partner.

\paragraph{RQ1 Summary}
\begin{itemize}
    \item Trust is highest with full explanations (Condition A) and lowest with no explanations (Condition C).
    \item Repeated Measures ANOVA reveals \textbf{a significant difference} in the mean trust scores across the three conditions ($p = 0.0006$).
    \item Post-hoc tests show that \textbf{Condition A results in significantly higher trust than Condition C}, while differences between A and B, and B and C, are not significant after correction.
    \item Item-level analysis shows that trust scores for \textbf{Q1, Q3, Q6, and Q7} vary significantly across conditions.
\end{itemize}

\subsection{RQ2: How Does the Level of Explanation Affect Developers’ Agreement with AI Recommendations?}

\paragraph{Descriptive statistics} 
Condition B achieves the highest AI agreement (89.22\%, 91/102), followed by Condition A (77.45\%, 79/102) and Condition C (66.67\%, 68/102).
The results show that most participants in Condition B (25/34) agreed with the AI across all three tasks, whereas in Condition A, agreement was split between partial and full agreement, and in Condition C, the fewest participants (7/34) agreed with the AI recommendation across all three tasks.

\paragraph{Statistical test}
A repeated-measures ANOVA reveals a significant main effect of condition on AI agreement, $F(2, 66) = 9.1042, p = 0.0003 < 0.05$. 

\paragraph{Post-hoc comparisons}
Post-hoc two-tailed paired t-tests (Table~\ref{tab:ttest_agreement}) show a significant difference between Condition B and Condition C, $t(33) = 4.49,\ p = 0.00008$, while the differences between Condition A and Condition B ($t(33) = -2.43,\ p = 0.0209$) and between Condition A and Condition C ($t(33) = 1.82,\ p = 0.0778$) are not statistically significant.
After Bonferroni correction for three comparisons ($\alpha = 0.017$), only B--C remains significant ($p_\text{corrected} = 0.00024 < 0.017$), indicating that participants agree with the AI significantly more often in Condition B than in Condition C. The other comparisons do not meet the corrected threshold ($p_\text{corrected} = 0.0628$ for A--B and $p_\text{corrected} = 0.2334$ for A--C).

\begin{table}[htbp]
\centering
\small
\caption{Post-hoc pairwise t-tests on AI agreement with Bonferroni correction}
\label{tab:ttest_agreement}
\begin{tabular}{cccc}
\toprule
Comparison & Raw $p$ & Corrected $p$ & Significant? \\
\midrule
A vs B     & 0.0209  & 0.0628         & No \\
A vs C     & 0.0778  & 0.2334         & No \\
{\color[HTML]{CB0000} \textbf{B vs C}} & {\color[HTML]{CB0000} \textbf{0.0001}} & {\color[HTML]{CB0000} \textbf{0.0002}} & {\color[HTML]{CB0000} \textbf{Yes}} \\
\bottomrule
\end{tabular}
\end{table}

\paragraph{Discussion}
These results indicate that the format and content of explanation can shape developers' willingness to agree with AI recommendations.
In particular, the more minimal feedback in Condition B is associated with the highest AI agreement.
One possible explanation is the review feedback provides just enough justification to make the recommendation seem reasonable, without directing attention to specific code fragments.
In contrast, Condition A combines review feedback and detailed explanation, yet participants may focus on the more salient inline content and pay less attention to the review comments.
It is important to note that greater agreement with AI recommendations is not necessarily a metric to optimize.
For example, adding more explanations to the AI recommendation may allow the user to form an opinion with more calibrated trust, leading to a more informed decision.
Users may overtrust Condition B, accepting its recommendations without sufficient validation.
Condition C provides only the AI recommendation, resulting in the lowest agreement overall.
Participants likely mistrust the system because it lacks a rationale and forces them to rely solely on their own judgment.
Notably, the AI recommendation always matches the original outcome of the pull request in the original project.
This means that lower agreement with AI recommendations also corresponds to lower accuracy in replicating the real-world outcome.
However, because we did not include recommendations where the AI recommendation didn't match the real-world outcome, we are unable to draw any conclusions about the match between AI recommendations and real-world outcomes.
Deliberately introducing flawed recommendations to study whether the explanations help developers detect them is a natural extension of this work, which we outline as future work in \Cref{sec:5-conclusion}.

The two trust signals we captured do not point in the same direction.
Self-reported trust (TXAI, RQ1) was the highest for Condition A (full explanations), whereas agreement behavior (accept/reject, RQ2) was the highest for Condition B (review feedback only).
We interpret this divergence as evidence of a gap between perceived and behavioral trust.
The explanation format that participants reported trusting the most (perception) was not the one that most influenced their actions (behavior).
This distinction is easy to overlook when trust is reported through a single lens, and we discuss its design implications in~\Cref{sec:implications}.
One possible explanation of this divergence is that more detailed explanations invite additional scrutiny by software engineers.
Condition A exposed more of the model's reasoning, giving participants more grounds on which to object to the recommendation.
This is consistent with the pattern we observed in our qualitative data (\Cref{sec:res-rq4}), where participants most commonly articulate disagreement with or uncertainty about the AI's rationale in Condition A.
Condition B may, in contrast, contain just enough justification to make the AI recommendation seem reasonable, without providing enough specific information for reviewers to contest it.
We offer this as an interpretation consistent with our qualitative design rather than a causal claim tested by our design.
Overall, the findings suggest a potential trade-off between trust and AI agreement that depends on the format and content of XAI support.

\paragraph{RQ2 Summary}
\begin{itemize}
    \item AI agreement is highest in Condition B (89.22\%), moderate in Condition A (77.45\%), and lowest in Condition C (66.67\%), indicating that the format of explanation influences participants’ alignment with AI recommendations.
    \item Repeated Measures ANOVA reveals \textbf{a significant difference} in the AI agreement across the three conditions.
    \item Post-hoc tests confirm that the AI agreement difference \textbf{between Condition B and C} is significant.
    \item Findings suggest that a more detailed explanation does not necessarily yield higher agreement, and concise feedback may support alignment better than an additional detailed explanation.
\end{itemize}

\subsection{RQ3: What Is the Impact of XAI on the Time Taken to Complete the Review Tasks?}

\paragraph{Descriptive statistics}
Condition A has an average time per task of \SI{166.42}{\second} (Md = \SI{130}{\second}, SD = \SI{153}{\second}, min = \SI{12}{\second}, max = \SI{1157}{\second}), Condition B \SI{147.75}{\second} (Md = \SI{124}{\second}, SD = \SI{99}{\second}, min = \SI{13}{\second}, max = \SI{419}{\second}), and Condition C \SI{164.18}{\second} (Md = \SI{132}{\second}, SD = \SI{147}{\second}, min = \SI{11}{\second}, max = \SI{1143}{\second}).
Overall, participants complete tasks slightly faster in Condition B than in Conditions A and C.

\paragraph{Statistical test}
A repeated-measures ANOVA indicates no significant difference in average completion time across conditions, $F(2, 66) = 0.5704, p = 0.5681 > 0.05$.

\paragraph{Discussion}
The analysis of task completion time across conditions reveals a trade-off between trust and efficiency.
Participants in Condition B spend the least amount of time on tasks, even though it does not yield the highest trust ratings.
In contrast, Condition A, which results in the highest trust scores, also requires the most time to complete.
The ANOVA test does not show a statistically significant difference in task duration across conditions.
Furthermore, potential confounding factors for this metric include the length of the output and participants' reading speed, as more explanations would require more reading.
For example, without XAI support, participants only see an accept/reject recommendation, whereas with XAI, they must read additional feedback or justifications before making a decision.
So while we did observe a nonsignificant difference between the conditions, we cannot come to a conclusion as to why this difference may have occurred.

\paragraph{RQ3 Summary}
\begin{itemize}
    \item Participants spend the least time per task in Condition B (\SI{147.75}{\second}), compared to Condition A (\SI{166.42}{\second}) and Condition C (\SI{164.18}{\second}).
    \item ANOVA results indicate \textbf{no significant effect} of condition on time ($p = 0.5681$).
    \item The format of XAI support does not significantly impact task duration.
\end{itemize}

\subsection{RQ4: What Reasons Do Software Engineers Provide for Accepting or Rejecting AI Code Review?}\label{sec:res-rq4}
We present our analysis based on the results of our qualitative coding process, as described in~\Cref{subsec:reason}.
In total, we coded 306 individual participant responses about why they accepted or rejected a code change.
Using a hierarchical codebook with paired positive and inverse codes, the three coders achieved acceptable inter-rater agreement on all reasoning responses (Krippendorff’s $\alpha = 0.7564$).

\begin{figure}[htbp]
  \centering
  \includegraphics[width=\linewidth]{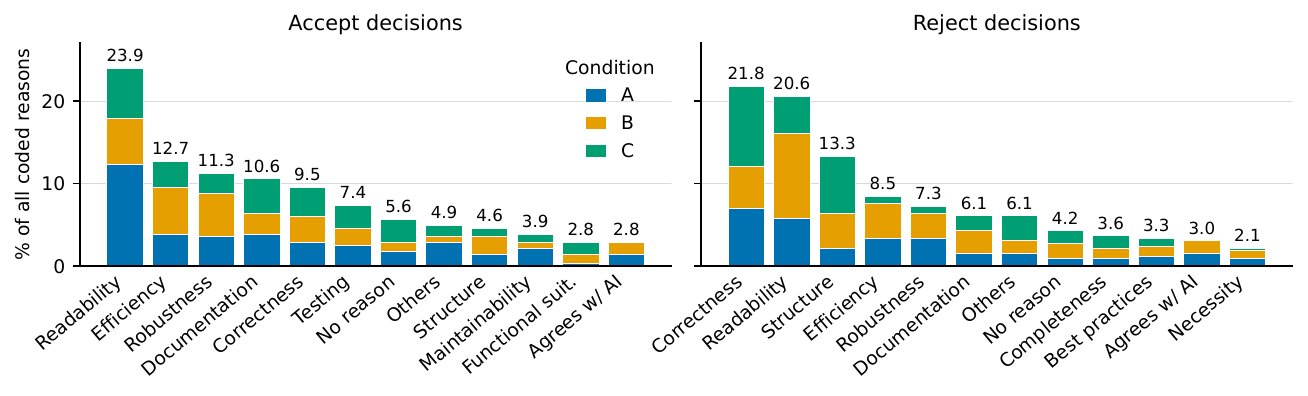}
  \caption{Frequency of codes assigned to responses, across the three conditions.}
  \Description{Two stacked bar charts showing the codes assigned to participant responses across the three conditions. Readability and correctness are the most frequent codes. The distribution across conditions is fairly even.}
  \label{fig:code_frequency}
\end{figure}

For the frequency distribution across all samples, we aggregated lower-level codes into higher-level groups.
Participants' reasons are largely related to maintainability, accounting for 55.37\% of all coded reasons in this category.
The most frequent subcategory is readability, cited both for accepting and rejecting changes, totaling 13.36\% of all reasons.
One Condition B response illustrates acceptance on these grounds: \textit{\enquote{The code is easily readable and the changes are solid, making the code more efficient.}}
Conversely, participants rejected changes perceived as harming readability, as in: \textit{\enquote{The code's naming is very inconsistent therefore affecting readability and overall output of the code.}}
AI-related reasons made up 3.42\% of responses; participants sometimes deferred explicitly to the AI's explanation, as in: \textit{\enquote{The updated code looks more efficient, and I'd follow the argumentation and recommendation of the AI.}}
\Cref{fig:code_frequency} shows the frequency distribution of codes by accept/reject decision and condition.
Readability is the most salient dimension in both directions: acceptance-driven readability gains ("2.4 Improved Code Readability," 23.94\% overall) cluster most heavily in Condition A, while rejection-driven readability losses ("I2.4 Reduced Readability," 20.61\% overall) cluster most heavily in Condition B.
Notably, "6.1 Agree with AI" never appeared in Condition C, in either direction.

\paragraph{Discussion}
When describing their reasoning, participants typically mention a bundle of improvements or regressions, as in this response from Condition A: \textit{\enquote{The changes significantly improve the code's robustness by adding error handling for empty lists and enhance readability and maintainability through clearer naming, type hints, and integrated doctest examples.}}
Correctness is the second-most dominant factor and the reason most often cited for overriding the AI, with participants pointing out specific faults that contradict the AI's recommendation.

Explicitly AI-related reasons are rare overall, but their distribution is informative.
All coded references to the AI recommendation occur in the two conditions with explanations: 11 in Condition A, 9 in Condition B, and none in Condition C.
When participants mention the AI, they frequently do so with reservations, as in \textit{\enquote{I am not entirely sure here, but kind of trust the AI in the end, in this situation, based mainly on a gut feeling. Although, I'd like to run the code and test it.}} (Condition A).
Every response coded as expressing disagreement or uncertainty with the AI's rationale came from Condition A, the condition with the most detailed explanations.
For example, one participant accepted a change the AI recommended rejecting: \textit{\enquote{I'm not sure, if I understand this one entirely. The argumentation seems plausible to me, but the final decision contradicts it. I'd stand by the proposed changes here.}}
Here, the explanation gave the participant enough of the model's reasoning to notice a discrepancy between the rationale and the recommendation, and they decided to act against the latter.
A reverse of this pattern can be observed with Condition C, where responses were most often coded as giving no specific reason at all: 13 in Condition C, 9 in Condition B, and 8 in Condition A.

Together, these results suggest that readability and correctness drive the accept/reject decision itself, while the presence and depth of explanations shape whether participants engage with the AI's reasoning at all.

\paragraph{RQ4 Summary}
\begin{itemize}
    \item Readability is the most common reason, with Condition A leaning toward accepting for better readability, Condition B rejecting for worse readability/efficiency, and Condition C rejecting for incorrect implementation.
    \item AI-related reasons account for only 3.42\% of all coded reasons.
    \item Agreement or disagreement with AI decisions was not mentioned for Condition C, where no explanations were shown.
\end{itemize}

\subsection{Implications for Practice}\label{sec:implications}
Based on our findings, we identify the following implications for software engineering practice:

\begin{itemize}
    \item \textbf{Include explanations to foster trust.}
    AI-based code review tools should provide explanations alongside recommendations.
    Condition C (no explanations) consistently showed the lowest trust scores, and participants never explicitly mentioned its recommendations as part of their reasoning, suggesting they may distrust or ignore unexplained recommendations.
    \item \textbf{Connect explanations directly within the code.}
    The inline explanations included in Condition A, which highlight specific code fragments, resulted in higher trust scores than auto-generated review comments alone.
    Designers of software tools should consider embedding explanations directly in the code diff rather than showing them separately from the text.
    \item \textbf{Be aware of the tradeoff between trust and agreement.}
    Increased trust does not straightforwardly translate into greater agreement: more detailed explanations raised trust, but also encouraged participants to scrutinize and ultimately override AI recommendations more often, challenging the assumption that more explanation uniformly yields both.
    Rather than indicating a negative outcome, such disagreement may reflect more calibrated trust and more informed decision-making. Designers should explicitly consider whether to optimize for user agreement or critical engagement.
    \item \textbf{Tailor the depth of explanations to the use case.}
    If the goal is rapid decision-making with high AI alignment, concise review feedback (Condition B) may suffice.
    If the goal is to support more deliberate, well-reasoned decisions, then richer explanations (Condition A) are preferable.
\end{itemize}

\subsection{Implications for Research}

\begin{itemize}
    \item \textbf{Report the level of explanation, not only performance.}
    As noted in~\Cref{sec:2-background}, work in this area is still evaluated predominantly by output quality, such as recommendation accuracy, while the transparency of those outputs remains underexplored~\cite{sun2022investigating}.
    Because the explanation format alone shifted both trust and agreement, while recommendation quality/correctness was held constant, performance-only evaluations omit a factor that measurably changes developer behavior.
    We argue that the level of explanation belongs among the variables that should be reported in studies evaluating AI recommendations.
    \item \textbf{Toward a theory of explanation detail, trust, and agreement.}
    We tentatively propose that trust increases with the level of detail in an explanation, whereas agreement does not, peaking instead at intermediate detail.
    We suggest scrutiny as the mechanism, since the detail that makes a recommendation credible also exposes more reasoning for a developer to contest (\Cref{sec:res-rq4}).
    As the differences between our two explained conditions did not reach significance after correction, we frame this as a hypothesis for future testing rather than an established relationship.
\end{itemize}

\subsection{Threats to Validity}\label{sec:dis-ttv}

\paragraph{External Validity:}
Although the code samples were drawn from a real-world repository, the study was conducted outside an actual project environment. 
Participants had no ownership of the code and faced no real consequences for accepting an incorrect recommendation. 
This may have affected both their trust ratings and accept/reject decisions, as developers may evaluate AI feedback more carefully when real project risks are involved. 
Therefore, our findings should be interpreted as evidence from a controlled study rather than as a direct representation of trust in professional practice.
However, we expect this effect to influence all three treatments equally.
The main objective of our study was to gain insights into the trust behavior of software developers when interacting with AI-assisted code reviews.
Therefore, we did not focus on creating an optimal XAI system for code reviews, as the actual design process was limited to mostly altering prompts for existing LLMs.
Interacting with a more sophisticated, purposefully designed XAI system may elicit different behaviors among software developers, thereby mitigating or amplifying the effects observed in this study.

We deliberately extracted all code change samples from a single repository (TheAlgorithms/Python) to minimize the influence of participants' prior domain knowledge and to reduce cognitive load during the tasks.
In addition, we examined only Python code and did not investigate whether the findings generalize to code written in other programming languages, which may differ in syntax, conventions, tooling, and typical development practices.
This choice limits the generalizability of our findings.
For instance, the application domain, such as banking software vs. e-commerce or internal tooling vs. consumer-facing software, may affect how AI explanations are generated and how developers perceive them.
Our results should therefore be read as evidence from one project context, and we encourage replication across heterogeneous repositories and programming languages.

\paragraph{Internal Validity:}
During the analysis of the experimental data, we noticed an error in the algorithm used to assign task sets to treatments.
Specifically, the task sets were shuffled by sorting them with a random comparator (Math.random() - 0.5).
This shorthand implementation does not produce a uniform distribution over permutations, making some assignments more likely than others.
A Fisher--Yates shuffle would have guaranteed equal probability across all permutations.
This led to some sample treatment combinations occurring less often than expected.
We cannot rule out the possibility that some of the observed between-condition differences reflect properties of the specific pull requests rather than the level of explanation alone.
Nevertheless, the design's core structure was preserved: each participant saw all three conditions, and within each condition, one easy, one medium, and one hard code change.
The full task distribution is documented in the replication package~\cite{gao_2026_21457282}, and we recommend that replications adopt a Fisher--Yates shuffle to ensure uniform assignment.

\paragraph{Construct Validity:}
When interpreting studies with proxy measures for mental constructs, one must always question whether the employed instruments accurately reflect the constructs.
We employed the trust questionnaire developed by Hoffman et al.~\cite{10.3389/fcomp.2023.1096257} to measure participants' perceived trust after using the systems during the experiment.
Measuring trust levels this way is a common approach, but it captures only a specific aspect of trust at a single point in time.
Hoffman et al. themselves recommend measuring trust at multiple points over a longer period of observation.

\section{Conclusion}
\label{sec:5-conclusion}
In this work, we investigated the impact of explainable AI on the trust software engineers place in AI-assisted code reviews.
Our goal was to address the lack of empirical evidence regarding XAI and code review, and to further analyze the relationship between XAI and human trust.
To achieve these goals, we conducted an online experiment with 34 participants.
Each participant was shown three conditions with varying levels of AI explanation and asked to accept or reject a code change.
Our results show that higher levels of explanation increase the degree of trust software engineers place in AI-enabled code review tools.
However, this trust did not correlate with their tendency to follow the AI's recommendation to accept or reject a code change.
Participants most often followed the AI recommendation for Condition B, which provided only moderate explanations.
Condition A, which offered the most elaborate explanations and elicited the highest level of perceived trust, led to more rejections of the AI system's recommendations, indicating a more complex relation between trust and agreement with AI decisions.
We did not find any statistically significant differences between the systems regarding the time taken to accept or reject a code change.
Code readability and correctness were the most commonly cited reasons for these decisions, and agreement with the AI recommendation was mentioned only in the conditions with explanations, never in Condition C.

Future work may expand upon this work and attempt to replicate its findings in a real-world case study.
Case studies on AI-assisted code review have already been conducted~\cite{sun2025bitsai, Aelsteinsson2025Rethinking, cihan2025automated}, but do not include XAI or trust measurements in their methodology.
We encourage future work to also shed light on and incorporate the effects that explainable AI could have in these cases.
Another natural extension of this work would be to study how the level of explanation interacts with the accuracy of the AI recommendation.
In the design of this study, we deliberately kept the quality of the recommendations constant by manually verifying their accuracy (see~\Cref{sec:models}).
This allowed us to isolate the effect of the explanation format on trust.
Evaluating whether explanations help developers detect incorrect recommendations would require a different design, such as a 2$\times$3 factorial crossing explanation level (none, review feedback only, full explanations) with recommendation correctness (correct, flawed), together with a correspondingly larger sample to preserve statistical power.
This direction could provide further insights towards mitigating overtrust in AI recommendations.

Our study captured trust in two ways: perceived trust via the self-reported TXAI scale and behavioral trust via participants' decisions to accept or reject the AI recommendation.
The measures were taken at a single point in time during an asynchronous online experiment.
Trust as a concept, however, is multi-faceted and future work could incorporate additional behavioral and process measures that are harder to obtain in this setting, for example, using eye-tracking tools to gather gaze information on individual sections of explanations, running think-aloud studies to understand how developers critically interrogate feedback before deciding, or using long-term telemetry data on how developers interact with explanations over longer time periods.
Such measures would allow a finer-grained calibration of the gap between perceived and behavioral trust observed in this study.

We compared participant subgroups by role, programming experience, and AI experience and found no consistent pattern indicating that particular explanation levels suit particular developer profiles.
However, this is a null result in our sample, not evidence of no effect.
Factors we did not control for, such as personality, domain familiarity, and project-level differences, may influence how explanations are perceived, and whether the level of explanation should be tailored to the developer or the project remains an open question worth dedicated study.
Given the results of this work, we emphasize the need for XAI in future studies of AI-assisted code review. Making the reasoning behind AI suggestions transparent to software engineers serves both to verify their trustworthiness and to inform the design of new tools.

\section{Data Availability}
We share all data and scripts produced as part of the study in our replication package~\cite{gao_2026_21457282}.
It includes the following artifacts:
\begin{itemize}
    \item The code change \textbf{samples} used in the experiment, including the AI recommendations and both levels of explanations.
    \item The \textbf{instruments} used to gather the experimental data, including the trust questionnaire and other questions asked to the participants.
    \item The \textbf{experimental data}, including all provided responses by participants regarding demographics, accept/reject decisions, timing of responses, and trust ratings.
    \item The Python \textbf{code} used to analyze the data, run the statistical tests, and generate visualizations.
    \item The \textbf{coding book} used to code the responses, the individual \textbf{codes} assigned by each of the three reviewers and the final agreed upon coding set for each individual participant response.
    \item The XML \textbf{templates} used to create the diagrams for the paper.
\end{itemize}

We omitted any personally identifiable information of the participants to protect their anonymity.

\bibliographystyle{ACM-Reference-Format}
\bibliography{bibliography}


\begin{thebibliography}{43}


\ifx \showCODEN    \undefined \def \showCODEN     #1{\unskip}     \fi
\ifx \showISBNx    \undefined \def \showISBNx     #1{\unskip}     \fi
\ifx \showISBNxiii \undefined \def \showISBNxiii  #1{\unskip}     \fi
\ifx \showISSN     \undefined \def \showISSN      #1{\unskip}     \fi
\ifx \showLCCN     \undefined \def \showLCCN      #1{\unskip}     \fi
\ifx \shownote     \undefined \def \shownote      #1{#1}          \fi
\ifx \showarticletitle \undefined \def \showarticletitle #1{#1}   \fi
\ifx \showURL      \undefined \def \showURL       {\relax}        \fi
\providecommand\bibfield[2]{#2}
\providecommand\bibinfo[2]{#2}
\providecommand\natexlab[1]{#1}
\providecommand\showeprint[2][]{arXiv:#2}

\bibitem[Atakishiyev et~al\mbox{.}(2025)]%
        {atakishiyev2025explainability}
\bibfield{author}{\bibinfo{person}{Shahin Atakishiyev}, \bibinfo{person}{Housam~KB Babiker}, \bibinfo{person}{Jiayi Dai}, \bibinfo{person}{Nawshad Farruque}, \bibinfo{person}{Teruaki Hayashi}, \bibinfo{person}{Nafisa~Sadaf Hriti}, \bibinfo{person}{Md~Abed Rahman}, \bibinfo{person}{Iain Smith}, \bibinfo{person}{Mi-Young Kim}, \bibinfo{person}{Osmar~R Za{\"\i}ane}, {et~al\mbox{.}}} \bibinfo{year}{2025}\natexlab{}.
\newblock \showarticletitle{Explainability of Large Language Models: Opportunities and Challenges toward Generating Trustworthy Explanations}.
\newblock \bibinfo{journal}{\emph{arXiv preprint arXiv:2510.17256}} (\bibinfo{year}{2025}).
\newblock
\href{https://doi.org/10.48550/arXiv.2510.17256}{doi:\nolinkurl{10.48550/arXiv.2510.17256}}


\bibitem[Atakishiyev et~al\mbox{.}(2024)]%
        {atakishiyev2024explainable}
\bibfield{author}{\bibinfo{person}{Shahin Atakishiyev}, \bibinfo{person}{Mohammad Salameh}, \bibinfo{person}{Hengshuai Yao}, {and} \bibinfo{person}{Randy Goebel}.} \bibinfo{year}{2024}\natexlab{}.
\newblock \showarticletitle{Explainable Artificial Intelligence for Autonomous Driving: A Comprehensive Overview and Field Guide for Future Research Directions}.
\newblock \bibinfo{journal}{\emph{IEEE Access}}  \bibinfo{volume}{12} (\bibinfo{date}{01} \bibinfo{year}{2024}), \bibinfo{pages}{101603--101625}.
\newblock
\href{https://doi.org/10.1109/ACCESS.2024.3431437}{doi:\nolinkurl{10.1109/ACCESS.2024.3431437}}


\bibitem[Aðalsteinsson et~al\mbox{.}(2025)]%
        {Aelsteinsson2025Rethinking}
\bibfield{author}{\bibinfo{person}{Fannar~Steinn Aðalsteinsson}, \bibinfo{person}{Björn~Borgar Magnússon}, \bibinfo{person}{Mislav Milicevic}, \bibinfo{person}{Adam~Nirving Davidsson}, {and} \bibinfo{person}{Chih-Hong Cheng}.} \bibinfo{year}{2025}\natexlab{}.
\newblock \showarticletitle{Rethinking Code Review Workflows with LLM Assistance: An Empirical Study}. In \bibinfo{booktitle}{\emph{2025 ACM/IEEE International Symposium on Empirical Software Engineering and Measurement (ESEM)}}. \bibinfo{pages}{488--497}.
\newblock
\href{https://doi.org/10.1109/ESEM64174.2025.00013}{doi:\nolinkurl{10.1109/ESEM64174.2025.00013}}


\bibitem[Baltes et~al\mbox{.}(2026)]%
        {baltes2025rethinking}
\bibfield{author}{\bibinfo{person}{Sebastian Baltes}, \bibinfo{person}{Timo Speith}, \bibinfo{person}{Brenda Chiteri}, \bibinfo{person}{Seyedmoein Mohsenimofidi}, \bibinfo{person}{Shalini Chakraborty}, {and} \bibinfo{person}{Daniel Buschek}.} \bibinfo{year}{2026}\natexlab{}.
\newblock \showarticletitle{{ On the Need to Rethink Trust in AI Assistants for Software Development: A Critical Review }}.
\newblock \bibinfo{journal}{\emph{IEEE Transactions on Software Engineering}} \bibinfo{volume}{52}, \bibinfo{number}{04} (\bibinfo{date}{April} \bibinfo{year}{2026}), \bibinfo{pages}{1265--1281}.
\newblock
\showISSN{1939-3520}
\href{https://doi.org/10.1109/TSE.2026.3659804}{doi:\nolinkurl{10.1109/TSE.2026.3659804}}


\bibitem[Bosu and Carver(2013)]%
        {bosu2013impact}
\bibfield{author}{\bibinfo{person}{Amiangshu Bosu} {and} \bibinfo{person}{Jeffrey~C Carver}.} \bibinfo{year}{2013}\natexlab{}.
\newblock \showarticletitle{Impact of peer code review on peer impression formation: A survey}. In \bibinfo{booktitle}{\emph{2013 ACM/IEEE International Symposium on Empirical Software Engineering and Measurement}}. IEEE, \bibinfo{pages}{133--142}.
\newblock
\href{https://doi.org/10.1109/ESEM.2013.23}{doi:\nolinkurl{10.1109/ESEM.2013.23}}


\bibitem[Chang et~al\mbox{.}(2024)]%
        {chang2024survey}
\bibfield{author}{\bibinfo{person}{Yupeng Chang}, \bibinfo{person}{Xu Wang}, \bibinfo{person}{Jindong Wang}, \bibinfo{person}{Yuan Wu}, \bibinfo{person}{Linyi Yang}, \bibinfo{person}{Kaijie Zhu}, \bibinfo{person}{Hao Chen}, \bibinfo{person}{Xiaoyuan Yi}, \bibinfo{person}{Cunxiang Wang}, \bibinfo{person}{Yidong Wang}, {et~al\mbox{.}}} \bibinfo{year}{2024}\natexlab{}.
\newblock \showarticletitle{A survey on evaluation of large language models}.
\newblock \bibinfo{journal}{\emph{ACM transactions on intelligent systems and technology}} \bibinfo{volume}{15}, \bibinfo{number}{3} (\bibinfo{year}{2024}), \bibinfo{pages}{1--45}.
\newblock
\href{https://doi.org/10.1145/3641289}{doi:\nolinkurl{10.1145/3641289}}


\bibitem[Cihan et~al\mbox{.}(2025a)]%
        {cihan2025automated}
\bibfield{author}{\bibinfo{person}{Umut Cihan}, \bibinfo{person}{Vahid Haratian}, \bibinfo{person}{Arda {\.I}{\c{c}}{\"o}z}, \bibinfo{person}{Mert~Kaan G{\"u}l}, \bibinfo{person}{{\"O}mercan Devran}, \bibinfo{person}{Emircan~Furkan Bayendur}, \bibinfo{person}{Baykal~Mehmet U{\c{c}}ar}, {and} \bibinfo{person}{Eray T{\"u}z{\"u}n}.} \bibinfo{year}{2025}\natexlab{a}.
\newblock \showarticletitle{Automated code review in practice}. In \bibinfo{booktitle}{\emph{2025 IEEE/ACM 47th International Conference on Software Engineering: Software Engineering in Practice (ICSE-SEIP)}}. IEEE, \bibinfo{pages}{425--436}.
\newblock
\href{https://doi.org/10.1109/ICSE-SEIP66354.2025.00043}{doi:\nolinkurl{10.1109/ICSE-SEIP66354.2025.00043}}


\bibitem[Cihan et~al\mbox{.}(2025b)]%
        {cihan2025evaluating}
\bibfield{author}{\bibinfo{person}{Umut Cihan}, \bibinfo{person}{Arda {\.I}{\c{c}}{\"o}z}, \bibinfo{person}{Vahid Haratian}, {and} \bibinfo{person}{Eray T{\"u}z{\"u}n}.} \bibinfo{year}{2025}\natexlab{b}.
\newblock \showarticletitle{Evaluating Large Language Models for Code Review}.
\newblock \bibinfo{journal}{\emph{arXiv preprint arXiv:2505.20206}} (\bibinfo{year}{2025}).
\newblock
\href{https://doi.org/10.48550/arXiv.2505.20206}{doi:\nolinkurl{10.48550/arXiv.2505.20206}}


\bibitem[Fan et~al\mbox{.}(2023)]%
        {fan2023large}
\bibfield{author}{\bibinfo{person}{Angela Fan}, \bibinfo{person}{Beliz Gokkaya}, \bibinfo{person}{Mark Harman}, \bibinfo{person}{Mitya Lyubarskiy}, \bibinfo{person}{Shubho Sengupta}, \bibinfo{person}{Shin Yoo}, {and} \bibinfo{person}{Jie~M Zhang}.} \bibinfo{year}{2023}\natexlab{}.
\newblock \showarticletitle{Large language models for software engineering: Survey and open problems}. In \bibinfo{booktitle}{\emph{2023 IEEE/ACM International Conference on Software Engineering: Future of Software Engineering (ICSE-FoSE)}}. IEEE, \bibinfo{pages}{31--53}.
\newblock
\href{https://doi.org/10.1109/ICSE-FoSE59343.2023.00008}{doi:\nolinkurl{10.1109/ICSE-FoSE59343.2023.00008}}


\bibitem[Faul et~al\mbox{.}(2009)]%
        {gpower}
\bibfield{author}{\bibinfo{person}{Franz Faul}, \bibinfo{person}{Edgar Erdfelder}, \bibinfo{person}{Axel Buchner}, {and} \bibinfo{person}{Albert-Georg Lang}.} \bibinfo{year}{2009}\natexlab{}.
\newblock \showarticletitle{Statistical power analyses using G*Power 3.1: Tests for correlation and regression analyses}.
\newblock \bibinfo{journal}{\emph{Behavior Research Methods}}  \bibinfo{volume}{41} (\bibinfo{date}{11} \bibinfo{year}{2009}), \bibinfo{pages}{1149--1160}.
\newblock
\href{https://doi.org/10.3758/BRM.41.4.1149}{doi:\nolinkurl{10.3758/BRM.41.4.1149}}


\bibitem[Faul et~al\mbox{.}(2007)]%
        {faul2007g}
\bibfield{author}{\bibinfo{person}{Franz Faul}, \bibinfo{person}{Edgar Erdfelder}, \bibinfo{person}{Albert-Georg Lang}, {and} \bibinfo{person}{Axel Buchner}.} \bibinfo{year}{2007}\natexlab{}.
\newblock \showarticletitle{G* Power 3: A flexible statistical power analysis program for the social, behavioral, and biomedical sciences}.
\newblock \bibinfo{journal}{\emph{Behavior research methods}} \bibinfo{volume}{39}, \bibinfo{number}{2} (\bibinfo{year}{2007}), \bibinfo{pages}{175--191}.
\newblock
\href{https://doi.org/10.3758/BF03193146}{doi:\nolinkurl{10.3758/BF03193146}}


\bibitem[Gao et~al\mbox{.}(2026)]%
        {gao_2026_21457282}
\bibfield{author}{\bibinfo{person}{Zhenhan Gao}, \bibinfo{person}{Marvin Muñoz~Barón}, \bibinfo{person}{Umm-e Habiba}, \bibinfo{person}{Daniel Graziotin}, {and} \bibinfo{person}{Stefan Wagner}.} \bibinfo{year}{2026}\natexlab{}.
\newblock \bibinfo{booktitle}{\emph{Evaluating the Impact of Explainable AI on Trust in AI-Assisted Code Review - Data, Code and Diagrams}}.
\newblock
\href{https://doi.org/10.5281/zenodo.21457282}{doi:\nolinkurl{10.5281/zenodo.21457282}}


\bibitem[{GitHub}(2023)]%
        {githubrestapi}
\bibfield{author}{\bibinfo{person}{{GitHub}}.} \bibinfo{year}{2023}\natexlab{}.
\newblock \bibinfo{title}{GitHub REST API documentation}.
\newblock \bibinfo{howpublished}{\url{https://docs.github.com/en/rest?apiVersion=2022-11-28}}.
\newblock


\bibitem[Gunning and Aha(2019)]%
        {gunning2019darpa}
\bibfield{author}{\bibinfo{person}{David Gunning} {and} \bibinfo{person}{David Aha}.} \bibinfo{year}{2019}\natexlab{}.
\newblock \showarticletitle{DARPA’s Explainable Artificial Intelligence (XAI) Program}.
\newblock \bibinfo{journal}{\emph{AI Magazine}}  \bibinfo{volume}{40} (\bibinfo{date}{06} \bibinfo{year}{2019}), \bibinfo{pages}{44--58}.
\newblock
\href{https://doi.org/10.1609/aimag.v40i2.2850}{doi:\nolinkurl{10.1609/aimag.v40i2.2850}}


\bibitem[Hellendoorn et~al\mbox{.}(2021)]%
        {hellendoorn2021towards}
\bibfield{author}{\bibinfo{person}{Vincent~J Hellendoorn}, \bibinfo{person}{Jason Tsay}, \bibinfo{person}{Manisha Mukherjee}, {and} \bibinfo{person}{Martin Hirzel}.} \bibinfo{year}{2021}\natexlab{}.
\newblock \showarticletitle{Towards automating code review at scale}. In \bibinfo{booktitle}{\emph{Proceedings of the 29th ACM Joint Meeting on European Software Engineering Conference and Symposium on the Foundations of Software Engineering}}. \bibinfo{pages}{1479--1482}.
\newblock
\href{https://doi.org/10.1145/3468264.3473134}{doi:\nolinkurl{10.1145/3468264.3473134}}


\bibitem[Hoffman et~al\mbox{.}(2023)]%
        {10.3389/fcomp.2023.1096257}
\bibfield{author}{\bibinfo{person}{Robert~R. Hoffman}, \bibinfo{person}{Shane~T. Mueller}, \bibinfo{person}{Gary Klein}, {and} \bibinfo{person}{Jordan Litman}.} \bibinfo{year}{2023}\natexlab{}.
\newblock \showarticletitle{Measures for explainable AI: Explanation goodness, user satisfaction, mental models, curiosity, trust, and human-AI performance}.
\newblock \bibinfo{journal}{\emph{Frontiers in Computer Science}}  \bibinfo{volume}{Volume 5 - 2023} (\bibinfo{year}{2023}).
\newblock
\showISSN{2624-9898}
\href{https://doi.org/10.3389/fcomp.2023.1096257}{doi:\nolinkurl{10.3389/fcomp.2023.1096257}}


\bibitem[Hossain et~al\mbox{.}(2025)]%
        {hossain2025explainable}
\bibfield{author}{\bibinfo{person}{Md~Imran Hossain}, \bibinfo{person}{Ghada Zamzmi}, \bibinfo{person}{Peter~R Mouton}, \bibinfo{person}{Md~Sirajus Salekin}, \bibinfo{person}{Yu Sun}, {and} \bibinfo{person}{Dmitry Goldgof}.} \bibinfo{year}{2025}\natexlab{}.
\newblock \showarticletitle{Explainable AI for medical data: Current methods, limitations, and future directions}.
\newblock \bibinfo{journal}{\emph{Comput. Surveys}} \bibinfo{volume}{57}, \bibinfo{number}{6} (\bibinfo{year}{2025}), \bibinfo{pages}{1--46}.
\newblock
\href{https://doi.org/10.1145/3637487}{doi:\nolinkurl{10.1145/3637487}}


\bibitem[Jacovi et~al\mbox{.}(2021)]%
        {jacovi2021formalizing}
\bibfield{author}{\bibinfo{person}{Alon Jacovi}, \bibinfo{person}{Ana Marasovi{\'c}}, \bibinfo{person}{Tim Miller}, {and} \bibinfo{person}{Yoav Goldberg}.} \bibinfo{year}{2021}\natexlab{}.
\newblock \showarticletitle{Formalizing trust in artificial intelligence: Prerequisites, causes and goals of human trust in AI}. In \bibinfo{booktitle}{\emph{Proceedings of the 2021 ACM conference on fairness, accountability, and transparency}}. \bibinfo{pages}{624--635}.
\newblock
\href{https://doi.org/10.1145/3442188.3445923}{doi:\nolinkurl{10.1145/3442188.3445923}}


\bibitem[Jian et~al\mbox{.}(2000)]%
        {jian2000foundations}
\bibfield{author}{\bibinfo{person}{Jiun-Yin Jian}, \bibinfo{person}{Ann Bisantz}, {and} \bibinfo{person}{Colin Drury}.} \bibinfo{year}{2000}\natexlab{}.
\newblock \showarticletitle{Foundations for an Empirically Determined Scale of Trust in Automated Systems}.
\newblock \bibinfo{journal}{\emph{International Journal of Cognitive Ergonomics}}  \bibinfo{volume}{4} (\bibinfo{date}{03} \bibinfo{year}{2000}), \bibinfo{pages}{53--71}.
\newblock
\href{https://doi.org/10.1207/S15327566IJCE0401_04}{doi:\nolinkurl{10.1207/S15327566IJCE0401_04}}


\bibitem[Krippendorff(2018)]%
        {krippendorff2018content}
\bibfield{author}{\bibinfo{person}{Klaus Krippendorff}.} \bibinfo{year}{2018}\natexlab{}.
\newblock \bibinfo{booktitle}{\emph{Content analysis: An introduction to its methodology}}.
\newblock \bibinfo{publisher}{Sage publications}.
\newblock


\bibitem[Lambiase et~al\mbox{.}(2025)]%
        {lambiase2025investigating}
\bibfield{author}{\bibinfo{person}{Stefano Lambiase}, \bibinfo{person}{Gemma Catolino}, \bibinfo{person}{Fabio Palomba}, \bibinfo{person}{Filomena Ferrucci}, {and} \bibinfo{person}{Daniel Russo}.} \bibinfo{year}{2025}\natexlab{}.
\newblock \showarticletitle{Investigating the role of cultural values in adopting large language models for software engineering}.
\newblock \bibinfo{journal}{\emph{ACM Transactions on Software Engineering and Methodology}} \bibinfo{volume}{35}, \bibinfo{number}{1} (\bibinfo{year}{2025}), \bibinfo{pages}{1--43}.
\newblock
\href{https://doi.org/10.1145/3725529}{doi:\nolinkurl{10.1145/3725529}}


\bibitem[Langenberg et~al\mbox{.}(2023)]%
        {langenberg2023tutorial}
\bibfield{author}{\bibinfo{person}{Benedikt Langenberg}, \bibinfo{person}{Markus Janczyk}, \bibinfo{person}{Valentin Koob}, \bibinfo{person}{Reinhold Kliegl}, {and} \bibinfo{person}{Axel Mayer}.} \bibinfo{year}{2023}\natexlab{}.
\newblock \showarticletitle{A tutorial on using the paired t test for power calculations in repeated measures ANOVA with interactions}.
\newblock \bibinfo{journal}{\emph{Behavior Research Methods}} \bibinfo{volume}{55}, \bibinfo{number}{5} (\bibinfo{year}{2023}), \bibinfo{pages}{2467--2484}.
\newblock
\href{https://doi.org/10.3758/s13428-022-01902-8}{doi:\nolinkurl{10.3758/s13428-022-01902-8}}


\bibitem[Li(2025)]%
        {githubranking}
\bibfield{author}{\bibinfo{person}{Evan Li}.} \bibinfo{year}{2025}\natexlab{}.
\newblock \bibinfo{title}{GitHub Ranking: Top 100 Python Repositories}.
\newblock \bibinfo{howpublished}{\url{https://github.com/EvanLi/Github-Ranking/blob/master/Top100/Python.md}}.
\newblock


\bibitem[Li et~al\mbox{.}(2019)]%
        {li2019deepreview}
\bibfield{author}{\bibinfo{person}{Heng-Yi Li}, \bibinfo{person}{Shu-Ting Shi}, \bibinfo{person}{Ferdian Thung}, \bibinfo{person}{Xuan Huo}, \bibinfo{person}{Bowen Xu}, \bibinfo{person}{Ming Li}, {and} \bibinfo{person}{David Lo}.} \bibinfo{year}{2019}\natexlab{}.
\newblock \showarticletitle{Deepreview: automatic code review using deep multi-instance learning}. In \bibinfo{booktitle}{\emph{Pacific-Asia Conference on Knowledge Discovery and Data Mining}}. Springer, \bibinfo{pages}{318--330}.
\newblock
\href{https://doi.org/10.1007/978-3-030-16145-3_25}{doi:\nolinkurl{10.1007/978-3-030-16145-3_25}}


\bibitem[Microsoft(2022)]%
        {codereviewer2022huggingface}
\bibfield{author}{\bibinfo{person}{Microsoft}.} \bibinfo{year}{2022}\natexlab{}.
\newblock \bibinfo{title}{CodeReviewer}.
\newblock \bibinfo{howpublished}{\url{https://huggingface.co/microsoft/codereviewer}}.
\newblock


\bibitem[Palikhe et~al\mbox{.}(2025)]%
        {Palikhe2025TowardsTA}
\bibfield{author}{\bibinfo{person}{Avash Palikhe}, \bibinfo{person}{Zhenyu Yu}, \bibinfo{person}{Zichong Wang}, {and} \bibinfo{person}{Wenbin Zhang}.} \bibinfo{year}{2025}\natexlab{}.
\newblock \showarticletitle{Towards Transparent AI: A Survey on Explainable Large Language Models}.
\newblock \bibinfo{journal}{\emph{ArXiv}}  \bibinfo{volume}{abs/2506.21812} (\bibinfo{year}{2025}).
\newblock
\href{https://doi.org/10.48550/arXiv.2506.21812}{doi:\nolinkurl{10.48550/arXiv.2506.21812}}


\bibitem[Perrig et~al\mbox{.}(2023)]%
        {perrig2023trust}
\bibfield{author}{\bibinfo{person}{Sebastian~AC Perrig}, \bibinfo{person}{Nicolas Scharowski}, {and} \bibinfo{person}{Florian Br{\"u}hlmann}.} \bibinfo{year}{2023}\natexlab{}.
\newblock \showarticletitle{Trust issues with trust scales: examining the psychometric quality of trust measures in the context of AI}. In \bibinfo{booktitle}{\emph{Extended abstracts of the 2023 CHI Conference on human factors in computing systems}}. \bibinfo{pages}{1--7}.
\newblock
\href{https://doi.org/10.1145/3544549.3585808}{doi:\nolinkurl{10.1145/3544549.3585808}}


\bibitem[Pornprasit and Tantithamthavorn(2024)]%
        {pornprasit2024fine}
\bibfield{author}{\bibinfo{person}{Chanathip Pornprasit} {and} \bibinfo{person}{Chakkrit Tantithamthavorn}.} \bibinfo{year}{2024}\natexlab{}.
\newblock \showarticletitle{Fine-tuning and prompt engineering for large language models-based code review automation}.
\newblock \bibinfo{journal}{\emph{Inf. Softw. Technol.}} \bibinfo{volume}{175}, \bibinfo{number}{C} (\bibinfo{date}{Nov.} \bibinfo{year}{2024}), \bibinfo{numpages}{12}~pages.
\newblock
\showISSN{0950-5849}
\href{https://doi.org/10.1016/j.infsof.2024.107523}{doi:\nolinkurl{10.1016/j.infsof.2024.107523}}


\bibitem[{Prolific}(2025)]%
        {prolific}
\bibfield{author}{\bibinfo{person}{{Prolific}}.} \bibinfo{year}{2025}\natexlab{}.
\newblock \bibinfo{title}{Prolific Participant Recruitment Platform}.
\newblock \bibinfo{howpublished}{\url{https://www.prolific.com/}}.
\newblock
\newblock
\shownote{Accessed: 2025-06-15}.


\bibitem[Saeed and Omlin(2023)]%
        {10.1016/j.knosys.2023.110273}
\bibfield{author}{\bibinfo{person}{Waddah Saeed} {and} \bibinfo{person}{Christian Omlin}.} \bibinfo{year}{2023}\natexlab{}.
\newblock \showarticletitle{Explainable AI (XAI): A systematic meta-survey of current challenges and future opportunities}.
\newblock \bibinfo{journal}{\emph{Know.-Based Syst.}} \bibinfo{volume}{263}, \bibinfo{number}{C} (\bibinfo{date}{March} \bibinfo{year}{2023}), \bibinfo{numpages}{24}~pages.
\newblock
\showISSN{0950-7051}
\href{https://doi.org/10.1016/j.knosys.2023.110273}{doi:\nolinkurl{10.1016/j.knosys.2023.110273}}


\bibitem[Said(2025)]%
        {said2025explaining}
\bibfield{author}{\bibinfo{person}{Alan Said}.} \bibinfo{year}{2025}\natexlab{}.
\newblock \showarticletitle{On explaining recommendations with Large Language Models: a review}.
\newblock \bibinfo{journal}{\emph{Frontiers in Big Data}}  \bibinfo{volume}{7} (\bibinfo{year}{2025}), \bibinfo{pages}{1505284}.
\newblock
\href{https://doi.org/10.3389/fdata.2024.1505284}{doi:\nolinkurl{10.3389/fdata.2024.1505284}}


\bibitem[Sarker et~al\mbox{.}(2023)]%
        {sarker2023toxispanse}
\bibfield{author}{\bibinfo{person}{Jaydeb Sarker}, \bibinfo{person}{Sayma Sultana}, \bibinfo{person}{Steven~R Wilson}, {and} \bibinfo{person}{Amiangshu Bosu}.} \bibinfo{year}{2023}\natexlab{}.
\newblock \showarticletitle{ToxiSpanSE: An explainable toxicity detection in code review comments}. In \bibinfo{booktitle}{\emph{2023 ACM/IEEE International Symposium on Empirical Software Engineering and Measurement (ESEM)}}. IEEE, \bibinfo{pages}{1--12}.
\newblock
\href{https://doi.org/10.1109/ESEM56168.2023.10304855}{doi:\nolinkurl{10.1109/ESEM56168.2023.10304855}}


\bibitem[Shi et~al\mbox{.}(2019)]%
        {shi2019automatic}
\bibfield{author}{\bibinfo{person}{Shu-Ting Shi}, \bibinfo{person}{Ming Li}, \bibinfo{person}{David Lo}, \bibinfo{person}{Ferdian Thung}, {and} \bibinfo{person}{Xuan Huo}.} \bibinfo{year}{2019}\natexlab{}.
\newblock \showarticletitle{Automatic code review by learning the revision of source code}. In \bibinfo{booktitle}{\emph{Proceedings of the AAAI Conference on Artificial Intelligence}}, Vol.~\bibinfo{volume}{33}. \bibinfo{pages}{4910--4917}.
\newblock
\href{https://doi.org/10.1609/aaai.v33i01.33014910}{doi:\nolinkurl{10.1609/aaai.v33i01.33014910}}


\bibitem[{Stack Overflow}(2025)]%
        {stackoverflow2025survey}
\bibfield{author}{\bibinfo{person}{{Stack Overflow}}.} \bibinfo{year}{2025}\natexlab{}.
\newblock \bibinfo{title}{2025 Stack Overflow Developer Survey}.
\newblock \bibinfo{howpublished}{\url{https://survey.stackoverflow.co/2025}}.
\newblock
\newblock
\shownote{Accessed: 2026-01-26}.


\bibitem[Sun et~al\mbox{.}(2022)]%
        {sun2022investigating}
\bibfield{author}{\bibinfo{person}{Jiao Sun}, \bibinfo{person}{Q~Vera Liao}, \bibinfo{person}{Michael Muller}, \bibinfo{person}{Mayank Agarwal}, \bibinfo{person}{Stephanie Houde}, \bibinfo{person}{Kartik Talamadupula}, {and} \bibinfo{person}{Justin~D Weisz}.} \bibinfo{year}{2022}\natexlab{}.
\newblock \showarticletitle{Investigating explainability of generative AI for code through scenario-based design}. In \bibinfo{booktitle}{\emph{Proceedings of the 27th International Conference on Intelligent User Interfaces}}. \bibinfo{pages}{212--228}.
\newblock
\href{https://doi.org/10.1145/3490099.3511119}{doi:\nolinkurl{10.1145/3490099.3511119}}


\bibitem[Sun et~al\mbox{.}(2025)]%
        {sun2025bitsai}
\bibfield{author}{\bibinfo{person}{Tao Sun}, \bibinfo{person}{Jian Xu}, \bibinfo{person}{Yuanpeng Li}, \bibinfo{person}{Zhao Yan}, \bibinfo{person}{Ge Zhang}, \bibinfo{person}{Lintao Xie}, \bibinfo{person}{Lu Geng}, \bibinfo{person}{Zheng Wang}, \bibinfo{person}{Yueyan Chen}, \bibinfo{person}{Qin Lin}, {et~al\mbox{.}}} \bibinfo{year}{2025}\natexlab{}.
\newblock \showarticletitle{Bitsai-cr: Automated code review via llm in practice}. In \bibinfo{booktitle}{\emph{Proceedings of the 33rd ACM International Conference on the Foundations of Software Engineering}}. \bibinfo{pages}{274--285}.
\newblock
\href{https://doi.org/10.1145/3696630.3728552}{doi:\nolinkurl{10.1145/3696630.3728552}}


\bibitem[{The Algorithms}(2025)]%
        {thealgorithms}
\bibfield{author}{\bibinfo{person}{{The Algorithms}}.} \bibinfo{year}{2025}\natexlab{}.
\newblock \bibinfo{title}{TheAlgorithms/Python}.
\newblock \bibinfo{howpublished}{\url{https://github.com/TheAlgorithms/Python}}.
\newblock


\bibitem[Tufano et~al\mbox{.}(2024)]%
        {tufano2024code}
\bibfield{author}{\bibinfo{person}{Rosalia Tufano}, \bibinfo{person}{Ozren Dabi{\'c}}, \bibinfo{person}{Antonio Mastropaolo}, \bibinfo{person}{Matteo Ciniselli}, {and} \bibinfo{person}{Gabriele Bavota}.} \bibinfo{year}{2024}\natexlab{}.
\newblock \showarticletitle{Code review automation: strengths and weaknesses of the state of the art}.
\newblock \bibinfo{journal}{\emph{IEEE Transactions on Software Engineering}} \bibinfo{volume}{50}, \bibinfo{number}{2} (\bibinfo{year}{2024}), \bibinfo{pages}{338--353}.
\newblock
\href{https://doi.org/10.1109/TSE.2023.3348172}{doi:\nolinkurl{10.1109/TSE.2023.3348172}}


\bibitem[Vaswani et~al\mbox{.}(2017)]%
        {vaswani2017attention}
\bibfield{author}{\bibinfo{person}{Ashish Vaswani}, \bibinfo{person}{Noam Shazeer}, \bibinfo{person}{Niki Parmar}, \bibinfo{person}{Jakob Uszkoreit}, \bibinfo{person}{Llion Jones}, \bibinfo{person}{Aidan~N. Gomez}, \bibinfo{person}{\L{}ukasz Kaiser}, {and} \bibinfo{person}{Illia Polosukhin}.} \bibinfo{year}{2017}\natexlab{}.
\newblock \showarticletitle{Attention is all you need}. In \bibinfo{booktitle}{\emph{Proceedings of the 31st International Conference on Neural Information Processing Systems}} (Long Beach, California, USA) \emph{(\bibinfo{series}{NIPS'17})}. \bibinfo{publisher}{Curran Associates Inc.}, \bibinfo{address}{Red Hook, NY, USA}, \bibinfo{pages}{6000–6010}.
\newblock
\showISBNx{9781510860964}


\bibitem[Watanabe et~al\mbox{.}(2024)]%
        {watanabe2024use}
\bibfield{author}{\bibinfo{person}{Miku Watanabe}, \bibinfo{person}{Yutaro Kashiwa}, \bibinfo{person}{Bin Lin}, \bibinfo{person}{Toshiki Hirao}, \bibinfo{person}{Ken'Ichi Yamaguchi}, {and} \bibinfo{person}{Hajimu Iida}.} \bibinfo{year}{2024}\natexlab{}.
\newblock \showarticletitle{On the use of chatgpt for code review: Do developers like reviews by chatgpt?}. In \bibinfo{booktitle}{\emph{Proceedings of the 28th International Conference on Evaluation and Assessment in Software Engineering}}. \bibinfo{pages}{375--380}.
\newblock
\href{https://doi.org/10.1145/3661167.3661183}{doi:\nolinkurl{10.1145/3661167.3661183}}


\bibitem[Widder et~al\mbox{.}(2021)]%
        {widder2021trust}
\bibfield{author}{\bibinfo{person}{David~Gray Widder}, \bibinfo{person}{Laura Dabbish}, \bibinfo{person}{James~D Herbsleb}, \bibinfo{person}{Alexandra Holloway}, {and} \bibinfo{person}{Scott Davidoff}.} \bibinfo{year}{2021}\natexlab{}.
\newblock \showarticletitle{Trust in collaborative automation in high stakes software engineering work: A case study at NASA}. In \bibinfo{booktitle}{\emph{Proceedings of the 2021 CHI Conference on Human Factors in Computing Systems}}. \bibinfo{pages}{1--13}.
\newblock
\href{https://doi.org/10.1145/3411764.3445650}{doi:\nolinkurl{10.1145/3411764.3445650}}


\bibitem[Wu et~al\mbox{.}(2026)]%
        {Wu2024}
\bibfield{author}{\bibinfo{person}{Xuansheng Wu}, \bibinfo{person}{Haiyan Zhao}, \bibinfo{person}{Yaochen Zhu}, \bibinfo{person}{Yucheng Shi}, \bibinfo{person}{Fan Yang}, \bibinfo{person}{Lijie Hu}, \bibinfo{person}{Tianming Liu}, \bibinfo{person}{Xiaoming Zhai}, \bibinfo{person}{Wenlin Yao}, \bibinfo{person}{Jundong Li}, \bibinfo{person}{Mengnan Du}, {and} \bibinfo{person}{Ninghao Liu}.} \bibinfo{year}{2026}\natexlab{}.
\newblock \showarticletitle{Usable XAI: 10 Strategies Towards Exploiting Explainability in the LLM Era}.
\newblock \bibinfo{journal}{\emph{ACM Trans. Knowl. Discov. Data}} \bibinfo{volume}{20}, \bibinfo{number}{6}, Article \bibinfo{articleno}{100} (\bibinfo{date}{July} \bibinfo{year}{2026}), \bibinfo{numpages}{57}~pages.
\newblock
\showISSN{1556-4681}
\href{https://doi.org/10.1145/3816150}{doi:\nolinkurl{10.1145/3816150}}


\bibitem[Yang et~al\mbox{.}(2023)]%
        {yang2023evacrc}
\bibfield{author}{\bibinfo{person}{Lanxin Yang}, \bibinfo{person}{Jinwei Xu}, \bibinfo{person}{Yifan Zhang}, \bibinfo{person}{He Zhang}, {and} \bibinfo{person}{Alberto Bacchelli}.} \bibinfo{year}{2023}\natexlab{}.
\newblock \showarticletitle{Evacrc: Evaluating code review comments}. In \bibinfo{booktitle}{\emph{Proceedings of the 31st ACM Joint European Software Engineering Conference and Symposium on the Foundations of Software Engineering}}. \bibinfo{pages}{275--287}.
\newblock
\href{https://doi.org/10.1145/3611643.3616245}{doi:\nolinkurl{10.1145/3611643.3616245}}


\end{thebibliography}

\end{document}